\documentclass[10pt,journal,compsoc]{IEEEtran}

\ifCLASSINFOpdf
\else
\fi

\pdfoutput=1

\usepackage{epsfig}
\PassOptionsToPackage{hyphens}{url}
\usepackage[hyperindex,breaklinks,hidelinks]{hyperref}
\usepackage[cmex10]{amsmath}
\usepackage{array}
\usepackage{mdwmath}
\usepackage{epsfig,hyperref, url, tabularx,color, graphicx, amsmath, dingbat, subfigure,amssymb}
\usepackage{multirow}
\usepackage{enumerate}
\usepackage{graphicx}
\usepackage{epstopdf}
\usepackage{booktabs}
\usepackage[numbers,sort&compress]{natbib}
\usepackage{algorithm}
\usepackage{algorithmicx}
\usepackage{algpseudocode}
\usepackage{bm}
\usepackage{xcolor}
\usepackage{url}

\begin{document}

\pagenumbering{arabic}
\date{}

\title{Nip in the Bud: Forecasting and Interpreting Post-exploitation Attacks in Real-time through Cyber Threat Intelligence Reports}

\author{Tiantian~Zhu,
        Jie Ying,
        Tieming Chen*,
        Chunlin Xiong,
        Wenrui Cheng,
        Qixuan Yuan,
        Aohan Zheng,
        Mingqi Lv,
        Yan Chen,~\IEEEmembership{Fellow,~IEEE}

\IEEEcompsocitemizethanks{
\IEEEcompsocthanksitem This work is supported in part by the following grants: National Natural Science Foundation of China under Grant No. 62002324, U22B2028 and U1936215. Fuxi Foundation of CCF-SANGFOR under Grant No. 20220201. Zhejiang Provincial Natural Science Foundation of China under Grant No. LQ21F020016 and LY20F020027. Key R\&D Projects in Zhejiang Province under Grant No. 2021C01117. "Ten Thousand People Program" Technology Innovation Leading Talent Project in Zhejiang Province under Grant No. 2020R52011; Major Program of Natural Science Foundation of Zhejiang Province under Grant No. LD22F020002.
\IEEEcompsocthanksitem T. Zhu, J. Ying, T. Chen*, W. Cheng, Q. Yuan, A. Zheng and M. Lv are with the College of Computer Science and Technology, Zhejiang University of Technology, Hangzhou 310023, China. E-mail: ttzhu@zjut.edu.cn, jieying@zjut.edu.cn, tmchen@zjut.edu.cn, zjutcwr@zjut.edu.cn, zjutyqx@zjut.edu.cn, zhengaohan198@163.com, mingqilv@zjut.edu.cn. *corresponding author
\IEEEcompsocthanksitem C. Xiong is with Department of Shenzhen Institutes of Advanced Technology, Chinese Academy of Sciences, Shenzhen 518052, China. E-mail: xiongchunlin@sangfor.com.cn.
\IEEEcompsocthanksitem Y. Chen is with Department of Electrical Engineering and Computer Science, Northwestern University, Evanston, IL 60208, USA. E-mail: ychen@northwestern.edu.
}
}


% \markboth{IEEE Transactions on Information Forensics and Security, ~Vol.~xx, No.~yy, Month~2020}%
% {Zhu \MakeLowercase{\textit{et al.}}: General, Efficient, and Real-time Data Compaction Strategy for APT Forensic Analysis}
\markboth{IEEE Transactions on Dependable and Secure Computing, ~Vol.~xx, No.~yy, Month~2024}%
{Zhu \MakeLowercase{\textit{et al.}}:Nip in the Bud: Forecasting and Interpreting Post-exploitation Attacks in Real-time through Cyber Threat Intelligence Reports}

\IEEEtitleabstractindextext{
\begin{abstract}
Advanced Persistent Threat (APT) attacks have caused significant damage worldwide. Various Endpoint Detection and Response (EDR) systems are deployed by enterprises to fight against potential threats. However, EDR suffers from high false positives. In order not to affect normal operations, analysts need to investigate and filter detection results before taking countermeasures, in which heavy manual labor and alarm fatigue cause analysts miss optimal response time, thereby leading to information leakage and destruction. Therefore, we propose Endpoint Forecasting and Interpreting (EFI), a real-time attack forecast and interpretation system, which can automatically predict next move during post-exploitation and explain it in technique-level, then dispatch strategies to EDR for advance reinforcement. First, we use Cyber Threat Intelligence (CTI) reports to extract the attack scene graph (ASG) that can be mapped to low-level system logs to strengthen attack samples. Second, we build a serialized graph forecast model, which is combined with the attack provenance graph (APG) provided by EDR to generate an attack forecast graph (AFG) to predict the next move. Finally, we utilize the attack template graph (ATG) and \textit{graph alignment plus algorithm} for technique-level interpretation to automatically dispatch strategies for EDR to reinforce system in advance. EFI can avoid the impact of existing EDR false positives, and can reduce the attack surface of system without affecting the normal operations. We collect a total of 3,484 CTI reports, generate 1,429 ASGs, label 8,000 sentences, tag 10,451 entities, and construct 256 ATGs. Experimental results on both DARPA Engagement and large scale CTI dataset show that the alignment score between the AFG predicted by EFI and the real attack graph is able to exceed 0.8, the forecast and interpretation precision of EFI can reach 91.8\%.

% 当前，高级持续威胁（APT）攻击在全球范围内呈现出高危高发态势。各类终端检测与响应系统（EDR）被企业部署以及时发现和应对潜在威胁。然而，EDR存在大量误报，为了不影响正常业务，分析人员必须人工对检测结果进行调查筛选之后才能采取对策，大量的人工劳动和告警疲劳使得分析人员错过了最佳响应时间，从而导致信息泄露与破坏发生。

% 为此，本文提出了EFI（Endpoint Forecasting and Interpreting）,一个攻击实时预测与解释系统，能够在攻击发生过程中自动化预测攻击者的下一步攻击手法，解释攻击技术并下发策略给EDR进行应急加固，帮助企业先攻击者一步，防患于未然。首先，我们利用大量CTI报告实现对可映射至系统底层日志的攻击实景图（ASG）的抽取，解决攻击样本不足问题；然后，我们构建序列化的图生成模型，并结合EDR产生的警报及攻击溯源图（APG）来生成攻击预测图（AFG），实现下一步post-exploitation attack的预测；最后，我们制作了攻击模板图（ATG），通过结合\textbf{图对齐算法+}对预测结果进行技术级别的解释，并能够自动化为EDR提供策略以提前加固。EFI能免受已有EDR误报的影响，能在不影响企业正常业务运行的情况下减少系统的攻击面。我们总计收集3483篇CTI报告，生成了1911份攻击图，人工绘制了256个技术模板，同时对EFI的各个模块进行了评估，其中技术模板的预测解释准确率能够达到91.8\%。

\end{abstract}

}

\maketitle
\IEEEraisesectionheading{\section{Introduction}\label{sec:intro}}

With the rapid development of the Internet, APT attacks have been frequently occurred around the world. 
% APT attacks are usually orchestrated by hacker groups with backgrounds in certain countries, governments, or other organizations, targeting core infrastructure (e.g., energy, transportation, and communication) and critical industries (e.g., military, finance, and medicine). Additionally, systems' confidentiality, integrity, and availability (CIA) can be severely compromised by these attacks. 
According to a report released by CSIS \cite{CSIS} in August 2023, there have been more than 7,000 significant APT attacks on governments, defense departments, and high-tech companies since 2006.

% To cope with the persistence and concealment of APT attacks, major global security vendors, such as Crowdstrike \cite{Crowdstrike} and Fortinet \cite{Fortinet}, have introduced a series of detection systems, e.g., Falcon \cite{Falcon}, and FortiGuard \cite{FortiGuard}, and have invested a large number of security practitioners in writing Cyber Threat Intelligence (CTI) reports to review and summarize complex cyber-attacks \cite{symantec_blog, microsoft_blog, talosintelligence_blog, crowstrike_blog}.

% To combat possible APT attacks in realistic operationss, firstly, enterprises would perform aforehand reinforcement at different levels such as applications, systems, infrastructure, and firmware, to lower security risks by potential attack points elimination and attack surface reduction of the system \cite{SystemsHardening}. Secondly, enterprises would deploy relevant attack detection systems to discover attacks promptly and make countermeasures accordingly. Meanwhile, a series of APT attack detection methods have been proposed in academia, and the mainstream is to use audit logs (e.g., ETW for Windows \cite{ETW}, and Audit for Linux \cite{audit}) to observe the interaction behavior between system entities for the sake of searching for attackers' traces. 
To combat possible APT attacks in the endpoint, EDR systems are deployed by enterprises to block malicious behaviors and provide suggestions for system repair. Audit logs (e.g., ETW for Windows \cite{ETW}, and Audit for Linux \cite{audit}) are often collected by EDR to observe the interaction behavior between system entities so as to search for traces of the attackers. 
% and the subsequent detection methods can be divided into two categories: \noindent \textbf{1) statistical-based and policy-based attack detection} \cite{hassan2019nodoze,xiong2020conan,hossain2020combating},  – relying on the idea of heuristic rules and tag propagation to determine the degree of abnormal behaviors. \noindent \textbf{2) learning-based attack detection} \cite{liu2019log2vec}, \cite{han2020unicorn} – based on modeling normal or abnormal system behavior, enabling the extraction of attacks by mapping the system behavioral states to some high-dimensional vectors and using classification or clustering algorithms. 
However, due to the high persistence of APT attacks \cite{zhu2021general}, endpoints will generate an enormous volume of system logs during the collection process, and majority of these logs result from normal applications \cite{hossain2017sleuth}. To find an attack is like looking for a needle in a haystack. In addition, some normal applications will perform like "suspicious behavior" (e.g., a newly installed benign browser may trigger six tactics in the ATT\&CK model \cite{attck} and raise an alarm \cite{milajerdi2019holmes}), which drowns true alarms in false positives. In real enterprise scenarios, response methods require human intervention limited by high false positive rate of detection (automated responses have a high probability of killing benign programs, which will undoubtedly affect normal operations). Therefore, it is arduous to reduce Mean-time-to-response (MTTR), which is one of the most crucial metrics for the defense: CrowStrike reports that the average lateral movement time for APT attacks is 1h 58mins \cite{lateral-movement}, which means that enterprises must complete a series of initiatives (e.g., detection, investigation, and response) within two hours to prevent information leakage or destruction.

We try to solve the above problems from a new perspective: Nip in the Bud. If we can get one step ahead of the attacker, automatically predict next move during post-exploitation, interpret the forecasting results in technique-level and dispatch the strategy to EDR for advance reinforcement, the above dilemmas will be addressed. By analyzing literature related to APT in the past five years, we find that: First, analysts often use the provenance graph in EDR to describe the activities of the attacker. Second, statistical-based and policy-based provenance graph analysis \cite{hassan2019nodoze,xiong2020conan,hossain2020combating,hossain2017sleuth} requires heavy manual labor and lacks generalization ability, while learning-based methods \cite{liu2019log2vec,han2020unicorn,chen2022apt} often face the problem of a scarcity of attack samples. Third, CTI reports written by professional security analysts incorporates knowledge of attack scenarios, TTPs of ATT\&CK model \cite{attck}, Indicators of Compromise (IOC), and their causal relationships, which provides detailed information on cyber attacks. Based on the above observations, in this paper, we attempt a best practice: strengthen attack (graph) samples with the help of CTI reports, and construct a learning-based model for post-exploitation attack forecast and interpretation. We summarize main challenges as follows:

\noindent \textbf{I: How to deal with the lack of samples in APT attack analysis?} Since the high complexity of APT attacks, it is time-consuming to simulate different attack techniques on real systems and collect corresponding logs.

% Since the high complexity of APT attacks, it is time-consuming to simulate different attack techniques on real systems and collect corresponding logs.

% First, due to the wide range of APT attacks, the attack logs available to different organizations are extremely limited, and attack-related samples are rarely shared for security and privacy reasons. Second, since the high complexity of APT attacks, it is time-consuming to simulate different attack techniques on real systems and collect corresponding logs. The above difficulties prevent defenders from gaining full insight into existing attack patterns.

\noindent \textbf{II: How to bridge the semantic gap between CTI reports described by natural language (high-level) and system logs (low-level)?} Current studies \cite{husari2017ttpdrill,satvat2021extractor,li2021attackg} have difficulty (e.g., the association of contextual semantics, the transformation of multi-hop equivalent semantic, and the extraction of dependency attributes) in automating the extraction of accurate ASG composed of entities and dependencies (events) with extremely different writing styles. 

% E.g., TTPDrill \cite{husari2017ttpdrill} can only extract node-level sequence information and cannot associate contextual semantics at the graph level, EXTRACTOR \cite{satvat2021extractor} extracts single attack nodes by default without considering the identity transformation of the attack process. AttackG \cite{li2021attackg} lacks the extraction of dependency attributes, and generates a large number of discrete sub-graphs due to the existence of redundant text.

\noindent \textbf{III: How to proactively and accurately predict post-exploitation attacks in real-time?} The judgment of the attacker's next move is overly depending on the experienced knowledge of security analysts, and the variety of attack techniques can make the decision-making difficult.

\noindent \textbf{IV: How to interpret the attack in technique-level based on the forecasting results?} In most cases, the attack forecasting results generated by learning-based models lack explanation in technique-level. Even with expertise, analysts are hard to directly use the forecasting results (graphs) to take countermeasures.

% To tackle the aforementioned challenges, in this paper, we extracts specific ASGs from massive CTI reports, combines cutting-edge EDR and attack forensic analysis technologies, and uses serialized graph generation networks to forecast attackers' attack paths. At the same time, we build technical templates according to the attack and defense knowledge base to interpret tactics and techniques of attacks, and finally construct an end-to-end system for automatic forecasting and interpreting post-exploitation attacks in a real-time manner.
% This system is able to help security analysts understand the attackers' next attack technologies and attack targets, so as to formulate protection countermeasures in advance. In summary, we make the following contributions:

To tackle the aforementioned challenges, we propose and open source EFI\footnote{https://github.com/EFI-Demo/Endpoint-Forecasting-and-Interpreting}, a real-time attack forecast and interpretation system. In summary, we make the following contributions:

\begin{itemize}

% \item{To the best of our knowledge, we are the first to work on automated forecasting and interpreting post-exploitation attacks on endpoints in real-time, which can get one step ahead of the attacker, automatically predict next move during post-exploitation, explain attacks in technique-level and dispatch strategies to EDR for advance reinforcement. It also can avoid the impact of existing EDR false positives, and reduce the attack surface of system without affecting the normal operations.}
\item{To the best of our knowledge, we are the first to work on automated forecasting and interpreting post-exploitation attacks on endpoints in real-time, which can avoid the impact of existing EDR false positives, and tremendously reduce the attack surface of system without affecting the normal operations.}

\item{For Challenges I and II, we build an ASG extraction module to achieve massive abstraction of ASGs (can be mapped to system-level logs to bridge the semantic gap) from CTI reports by adopting heuristic natural language processing (NLP) pipeline.}

\item{For Challenge III, we construct an AFG generation module with the help of a serialized graph forecast model for realizing the sub-graph prediction. This module simultaneously captures node attributes, edge attributes, and edge time ordering of the graph to maximize the forecasting fidelity of post-exploitation attacks.}

\item{For Challenge IV, we construct ATGs based on the atomic red team technique \cite{atomic-red-team}, and incorporate an innovative \textit{graph alignment plus algorithm} to provide further interpretability for AFG, which facilitates automation to graph investigation for EDR to implement advance reinforcement.}

\item{We conduct a detailed evaluation of all the modules in EFI. The results show that EFI can generate an AFG within 5s, interpret the AFG in technique-level within 5mins and obtain an alignment score of more than 0.8, the forecast and interpretation precision of EFI can reach 91.8\%.}

\end{itemize}

% The rest of this article is organized as follows: we present the background and motivation in Section~\ref{sec:motivation}. Section~\ref{sec:intro} introduces the overview of our system. Section~\ref{sec:sysdesign} describes the architecture of our system. Section~\ref{sec:implement} proposes our implementation approaches. Section~\ref{sec:evaluation} shows the evaluation of the system. Section~\ref{sec:discuss} discusses the proposed approaches and analyzes their limitations. Section~\ref{sec:related} provides the related work, and Section~\ref{sec:conclusion} concludes our work.

\section{Motivation and Background}\label{sec:motivation}

\subsection{Motivating Example}\label{sec:2_1}

We assume there is a real-world campaign that the group APT28 \cite{apt28-hacking} is trying to compromise the enterprise system to steal confidential data. Refer to a known set of tactics and techniques that APT28 commonly use listed by MITRE \cite{apt28-mitre}. The attacker of APT28 induces the victim to download malware, then executes the malicious file , collects the target data, and finally steals confidential data over C2 Channel. Meanwhile, an employee downloads and installs the firefox browser locally, and then firefox collects local data for configuration. 
% Obviously, the two examples have great similarity in procedure and both hit the alert condition (most of EDRs typically detect attacks based on predefined rules \cite{edr_market, hassan2020tactical, hassan2020omegalog,hossain2017sleuth,milajerdi2019holmes} or anomalous scores \cite{liu2019log2vec, han2020unicorn}).
%The behavior of firefox hits the alert condition (most of EDRs typically detect attacks based on predefined rules \cite{edr_market, hassan2020tactical, hassan2020omegalog} or anomalous scores \cite{liu2019log2vec, han2020unicorn}) and raises an alarm (false positive). 

\par
\textbf{Limitations of EDR tools}: Existing EDR tools have a high rate of false positives, the two examples have great similarity in procedure and both hit the rule based alert condition "download \& execution" \cite{edr_market, hassan2020tactical,hassan2020omegalog,hossain2017sleuth,milajerdi2019holmes}. While for anomaly based methods \cite{liu2019log2vec, han2020unicorn}, the aforementioned situation becomes even more challenging to distinguish. More specifically, assuming that $ABCD$ represents a sequence of actions performed by a legitimate program, and $ABCX$ represents a sequence of actions carried out by an attacker. Although detecting the presence of sensitive behavior in the system solely based on $ABC$ can trigger an alert from EDR, it is not sufficient to determine whether an attack is taking place (i.e., further analysis is needed to determine if action $D$ or $X$ follows). By the time the alert is triggered after the attacker has completed $ABCX$, it is already too late for EDR to response as the damage has already occurred. To avoid the \textbf{choice dilemma} (i.e., to kill or not to kill when alerts like $ABC$ are triggered), it is necessary for analysts to proactively forecast the attacker's next move in real-time and carry out targeted defensive measures before the ultimate goal is achieved during post-exploitation.

\par
Therefore, we propose EFI as a third-party tool for existing EDRs, capable of forecasting next move during post-exploitation, explaining attacks in technique-level and dispatching strategies to EDR for advance reinforcement. \textbf{After reinforcement, we consider two status}: 1) There exists a false alarm and the process does not perform any suspicious actions (i.e., $D$), then the restriction can be removed heuristically (e.g., exceeding average lateral movement time \cite{lateral-movement}). 2) If the process performs the predicted operation (i.e., $X$), the EDR will block the event, raise a high-level alarm and notify the analyst. Since the corresponding point has been reinforced in advance, the attacker will not achieve his goal. \textbf{It is important to note that in order to ensure the accuracy of the forecast results, the EDR can automatically verify whether the predicted event, combined with the existing behavior, satisfies a complete context of an attack (i.e., graph investigation). If it does, EDR systems will directly block the corresponding predicted event} (also mentioned in Section~\ref{sec:4_4}).

\subsection{Causal Graph and Heterogeneous Graph}
\par
\textbf{Causal graph} is a data structure extracted from low-level system logs representing causal relationships (e.g., exec, read, etc.) between subjects (e.g., processes) and objects (e.g., files). 
% Causal graph consists of nodes representing subjects and objects, and edges representing actions between subjects and objects . 
We here consider a directed causal graph whose edges point from a subject to an object.
\par
\textbf{Heterogeneous graph} is defined as G = (V, E), the set of nodes is $ V = (v_{1},...,v_{n}) $ and the set of edges between nodes is $ E = ({v_{i}, v_{j} |v_{i}, v_{j} \in V}) $, where each node and edge corresponds to a specific attribute. Denoting the set of node attributes by $\Gamma _{v} $ and the set of edge attributes by $\Gamma _{e} $, the heterogeneous graph requires $\mid \Gamma _{v} \mid + \mid \Gamma_{e} \mid > 2$. Obviously, causal graph is a kind of heterogeneous graph, and all the graph structures mentioned in this paper are heterogeneous graphs.

\section{Overview}\label{sec:overview}

\subsection{Graph Description}\label{sec:3_1}
\par
% A variety of graphs exist in this paper. For ease of reference and to reflect their differences in purpose and origin, we divide them into four categories and describe them separately.
In this paper, we divide attack graphs into four categories as follows: \textbf{Attack scene graph (ASG)} is automatically extracted from CTI reports for training graph forecast model. It can be mapped to low-level system logs to strengthen attack samples. \textbf{Attack provenance graph (APG)} %is the sub-graph part that EDR extracts from the causal graph related to this initial detection point (IDP). This part of the work is also called forensic analysis, the specific operations of which are not covered in this paper.
is a potential attack chain extracted by EDR based on alarm point  through forensic analysis from low-level system logs and used for subsequent post-exploitation attacks forecasting. \textbf{Attack forecast graph (AFG)} is generated by the graph forecast model, which contains the initial APG and next operation(s) of potential attackers. \textbf{Attack template graph (ATG)} 
%also called technical templates, is an atomic graph that we manually construct for interpreting AFG, based on the description of atomic techniques and example code.
is a technique template that we construct from atomic red team \cite{atomic-red-team} for interpreting AFG.
% It should be noted that these four graphs, although different in purpose and origin, are all consistent in level, i.e., they are all heterogeneous graphs that can be mapped to the underlying system logs. Moreover, ASG, APG and ATG are concrete, i.e., the nodes all have naming, while AFG is abstract, i.e., the nodes have only types without naming. We therefore propose the \textit{graph alignment plus algorithm} applied to abstract graphs to interpret AFG using ATG to automate policy dispatching for reinforcement in advance.

\subsection{Threat Model}
\label{sec:3_2}
\par
% In this paper, we consider a complex and sophisticated enterprise environment that is currently the target of choice for APT attackers, and the APT attack is highly destructive \cite{apt-kill-chain}, which can cause huge economic losses. First, we assume that the enterprise has deployed a mature EDR tool in its production environment, but the EDR has a high false positive problem, and the enterprise has high requirements for the stability of its production environment and cannot directly kill suspicious processes based on alerts. Next, we assume that the APT attack occurs after the EDR starts monitoring the victim hosts, and that the EDR tool is not compromised and can properly issue alerts, capture IDPs, and provide a attack provenance graph of this alert in real time. Finally, hardware Trojan, side channel and backdoor attacks are not considered in this paper.

In this paper, we consider a complex enterprise environment. We assume that the enterprise has deployed a mature EDR system with massive false positives (a common problem of EDR in the industry). The enterprise has high requirements for the stability of its production environment and cannot directly kill all processes related to alarms. Next, we assume that the APT attack occurs after the EDR is deployed, the EDR can issue alarms without being compromised, and provide the APG of the alarm in real-time (e.g., the method proposed by \cite{xiong2020conan}). Finally, we do not consider attack methods that cannot be detected by EDRs, such as hardware Trojans, side channels, and backdoors.

% \subsection{Design Goals}
% The design goals of EFI are as follows:
% \par
% \textbf{G1: Extract attack scene graphs.}
% %The system should be able to automate the processing of a large number of unstructured CTI reports and the construction of attack scene graphs that can be mapped to the low-level system logs.
% EFI needs to process a large number of unstructured CTI Reports and automatically extract ASGs that can be mapped to the low-level system logs.
% \par
% \textbf{G2: Generate attack forecast graphs.} EFI should be able to generate AFGs that represent the future actions of the attacker in real-time based on the APGs provided by EDR.
% \par
% \textbf{G3: Interpret the prediction results.} EFI should be able to interpret the prediction results with ATGs at the technique-level and dispatch the strategy to EDR for emergency reinforcement.

\subsection{Our Approach}
\label{sec:3_3}
\par
\begin{figure*}[t]
% \caption{The architecture of EFI. EFI extracts ASGs from a large number of open source CTI reports and trains the graph forecast model. The APG provided by EDR is then fed into model to predict the AFG, which is then interpreted using the \textit{graph alignment plus algorithm} and ATGs. Finally, the interpretation results are used to dispatch strategies to EDR for advance reinforcement.}
%\label{fig:System Architecture}
\centering
\includegraphics[scale=0.15]{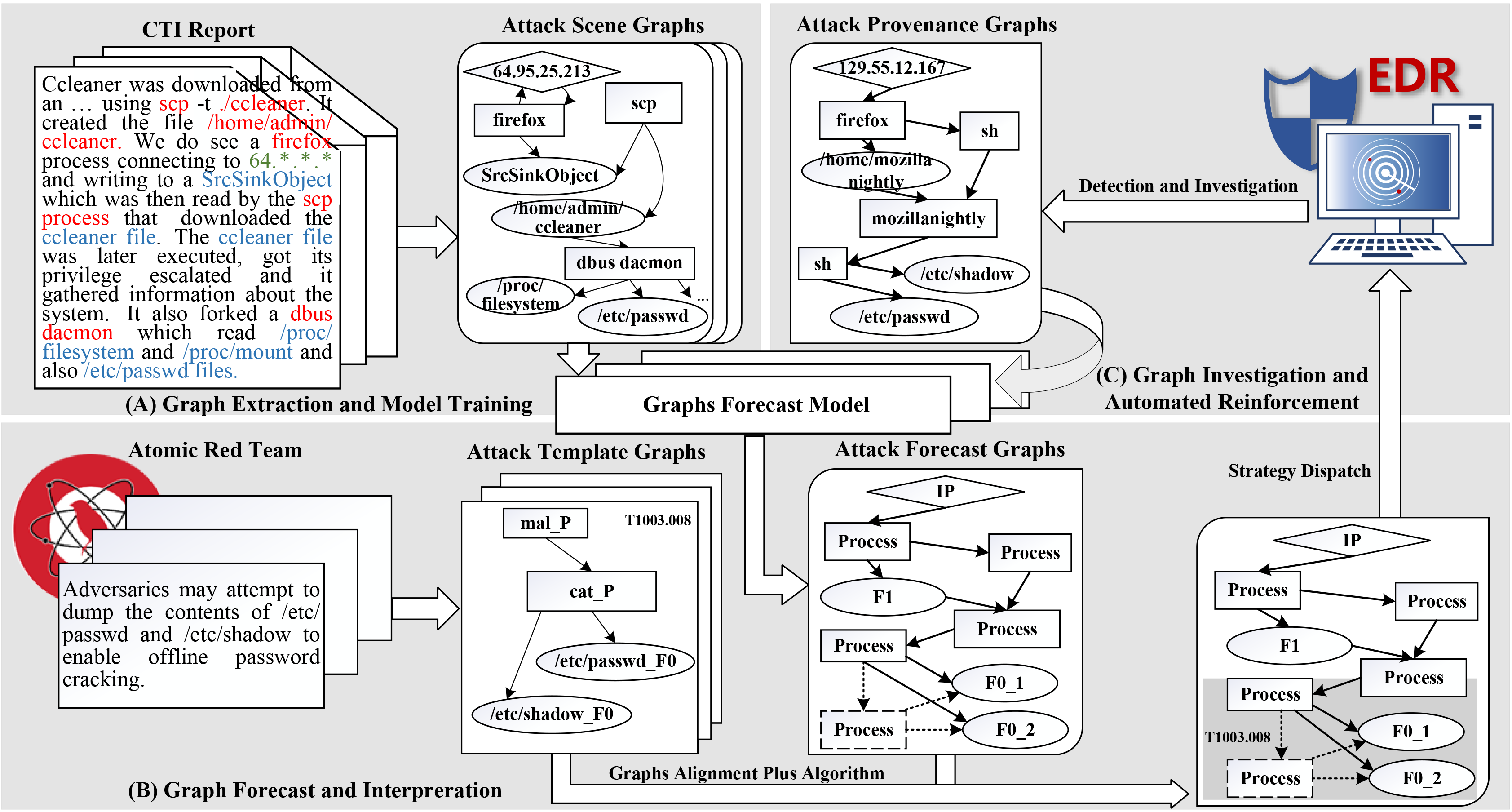}
\caption{The architecture of EFI. EFI extracts ASGs from open source CTI reports and trains the graph forecast model.   The APG provided by EDR is then fed into the model to predict the AFG, which is then interpreted using the \textit{graph alignment plus algorithm} and ATGs. Finally, the interpretation results are used to dispatch strategies to EDR for advance reinforcement.}
\label{fig:System_Architecture}
% \vspace{-0.2cm}
\end{figure*}

The architecture of EFI is shown in Figure~\ref{fig:System_Architecture}. First, EFI automates the extraction of ASGs (contain entities and events that can be mapped to low-level system logs) from unstructured CTI reports without human intervention. Figure~\ref{fig:asg_diffsys_compare} gives an example of ASG from the Darpa Engagement CTI report.
The attributes of system entities and events are shown in Table~\ref{table:system_entities} and Table~\ref{table:system_event}, respectively. The reason we divide file entities into four attributes is that different file types imply different purposes of attackers. Fine-grained division can also effectively reduce false positives of subsequent AFG interpretations.
% Finally, we add category suffixes to entities in the ASG for real-time distinction, while the graph contains only entities related to the attack behavior and no other information (the main goal is a complete attack graph and simplicity). 
\par
% We then use ASGs to train a serialized graph forecast model that maximizes the learning of the graph distribution in them from both the node and edge level. In addition, we also perform targeted optimization of the graph forecast model to capture features of the graph such as node type, edge type, and overall structure simultaneously. The trained model takes the attack provenance graph (provided by the deployed EDR tools) as input and serially predicts the type of the next node and the dependencies (existence and specific type) between that node and existing nodes to finally obtain the attack forecast graphs (Goal G2).

Then, EFI uses ASGs to train a serialized graph forecast model that maximally learns the graph distribution. In addition, EFI also performs targeted optimization of the graph forecast model to capture the node attributes, edge attributes and edge time ordering of the graph at the same time. The trained model takes the APG (provided by the deployed EDR system) as input to serially forecasts: 1) the attribute of the next node, and 2) the dependency between the new node and all the existing nodes (whether there is an edge and the type of the specific edge), to finally obtain the AFG.
% (Goal G2)

\par
% Finally, we interpret the obtained AFG. Specifically, we construct the corresponding attack template graph based on the atomic red team technique \cite{atomic-red-team}, as shown in the appendix \ref{fig:technique template schematic}, i.e., the atomic attack described as a graph at the system logs level, which we open source \footnote{open source}.We then implemented \textit{graph alignment plus algorithm}, which calculates the similarity between the AFG and different ATGs, and uses ATGs that exceed the threshold as the technique aspect of the interpretation of the AFG (Goal G3). Then, EFI will use the interpretation results to dispatch policies to EDR for advance reinforcement, thus avoiding the impact of EDR false positives and being able to reduce the attack surface of the system without affecting the normal business operation of the enterprise.
% Detailed system design and implementation details will be given in the next section, and only an overview of the approach is given here.
Besides, we construct ATGs based on the atomic red team technique \cite{atomic-red-team} to interpret the AFG (Details of ATGs are shown in Section~\ref{subsec:afg_interpretation} and Appendix~\ref{sec:appendix_atgexam}, and we open source ATGs). We propose the \textit{graph alignment plus algorithm} to calculate the similarity between the AFG and different ATGs, and use the ATGs exceeding the threshold as the technical explanation of AFG. 

Finally, EFI will combine the interpretation results and technique-level attack detection/investigation tools (e.g., APTSHIELD \cite{zhu2023aptshield}) to dispatch strategies to EDR for advance reinforcement, thus avoiding the impact of EDR false positives and being able to reduce the attack surface of the system without affecting the normal operations. It is important to note that whether it is the ASG, AFG, APG or ATG, \textbf{EFI only retains its node attributes and discards the node names}. This is because the latter is susceptible to dynamic changes by attackers and EFI prefers to learn the graph distribution representing the attack procedure.

\begin{table}[t]
\caption{The attribute of system entities in EFI. The second column indicates the symbol of entity attribute in the system.}
\setlength{\abovecaptionskip}{2cm}
\label{table:system_entities}
\centering
\begin{tabular}{|c|c|c|}
\hline
\textbf{Entity}                & \textbf{Symbols} & \textbf{Description}                                                                 \\ \hline
\multirow{5}{*}{\textbf{File}} & F0               & \begin{tabular}[c]{@{}c@{}}Sensitive files \\ such as '/etc/passwd'.\end{tabular}    \\ \cline{2-3} 
                               & F1               & \begin{tabular}[c]{@{}c@{}}Library files\\ such as '*.dll', '*.so'.\end{tabular}     \\ \cline{2-3} 
                               & F2               & \begin{tabular}[c]{@{}c@{}}Executable files\\ such as '*.exe', '*.vbs'.\end{tabular} \\ \cline{2-3} 
                               & F3               & Other files.                                                                         \\ \cline{2-3} 
                               & FR                & \begin{tabular}[c]{@{}c@{}}Registry files\\ such as 'HKCR', 'HKLM'\end{tabular}      \\ \hline
\textbf{Process}               & P                & Such as PID, user, group.                                                            \\ \hline
\textbf{Socket}                & S                & Such as IP, Domain.                                                                  \\ \hline
\end{tabular}
\end{table}

\begin{table}[ht]
\setlength{\abovecaptionskip}{-0cm}
\caption{The attribute of system events in EFI. P indicates process, F indicates file, and S indicates socket in column 2.}
\label{table:system_event}
\centering
\begin{tabular}{|c|c|c|c|}
\hline
\textbf{Type}                                         & \textbf{Sub.} & \textbf{Obj.} & \textbf{Description}                                                                           \\ \hline
Write                                                 & P                & F               & A process writes a file.                                                                       \\ \hline
Execute                                                & P                & F               & A process executes a file                                                                       \\ \hline
Read                                                  & F                & P               & A file is read by a process.                                                                   \\ \hline
Send                                                  & P                & S               & \begin{tabular}[c]{@{}c@{}}A process sends information to \\ a socket.\end{tabular}            \\ \hline
Receive                                               & S                & P               & \begin{tabular}[c]{@{}c@{}}Information from a socket is \\ received by a process.\end{tabular} \\ \hline
\begin{tabular}[c]{@{}c@{}}Frok/\\ Clone\end{tabular} & P                & P               & A process starts a new process.                                                                \\ \hline
\end{tabular}
\vspace{-0.3cm}
\end{table}
\section{System Design}\label{sec:sysdesign}
% In this section, we will introduce key components in EFI, including  ASG extraction, AFG Forecasting, and AFG Interpretation.
\subsection{ASG Extraction}
\label{sec:4_1}
\par 
In this section, we present the details about extracting ASGs that can be mapped to the low-level system logs from unstructured CTI reports. A useful CTI report would provide an exhaustive chronological description of the attacker's actions on the victim host. Normally, these descriptions include the involved entities as well as their interactions. To extract the ASGs, it is imperative to bridge the huge semantic gap between unstructured CTI and low-level system logs. 
% Although some related works have been done by iACE \cite{liao2016acing}, AttackG \cite{li2021attackg}, and EXTRACTOR \cite{satvat2021extractor}, the problems of text redundancy, missing attack-related entities, and insufficient dependency still exist. 
We summarize several challenges as follows:

\par
\textbf{C1 Text Verbosity.}
In general, CTI reports are redundant. Sentences that related to the attack are often drowned out in the tedious text. For example, only 9 sentences describe the attack out of more than 300 lines of reports on Flamer \cite{3rd-update-flamer}. %In addition, after traversing 3,483 CTI reports, we found that the average length of attack-related description per report was only 11 sentences.
\par
\textbf{C2 Related Entity Recognition.}
% Entity Recognition (ER) is a common problem in the field of NLP. After analyzing a large number of CTI reports, we found it necessary to correctly identify two types of entities: IOC and non-IOC entities. To be specific, IOC entities refer to domain-specific terms, including IP addresses, paths, and registries. Non-IOC entities are components that refer to IOC entities by various expressions, such as Process/Attacker for malicious processes like *.exe. Besides, it is essential to address the ubiquity problem of reference, which is classified as anaphora and co-reference. To illustrate, "it" and "the user" are both used to refer to Host A, as shown in Figure \ref{fig:asg_diffsys_compare}.
We divide attack-related entities into IOC and non-IOC entities in CTI reports. IOC entities refer to domain-specific terms, including IP addresses, paths, and registries. Non-IOC entities are components that represent important entities by various expressions (e.g., using a group name for a subject to describe an attack scenario), which can cause semantic deficiencies or confusion in the ASG if they are not properly recognized. Besides, it is essential to address the ubiquity problem of reference, which is classified as anaphora and co-reference. To illustrate, "it" and "the user" are both used to represent Host A, as shown in the text of Figure~\ref{fig:asg_diffsys_compare}.
\par
\textbf{C3 Dependency Extraction.}
% To properly construct the ASG, we extract the relationships between different entities after identifying them, these relationships can be mapped to the low-level system logs (e.g. read, write, fork, send, etc.).  However, CTI reports often use multiple ways to describe the relationships between entities, such as shortcuts (e.g. cat, ifconfig, etc. in Figure \ref{fig:asg_diffsys_compare}), which can affect the use of NLP tools and lead to inaccurate extraction of dependencies.
To properly construct the ASG, we need to extract relationships between different attack entities. However, even for the same attack, different analysts produce heterogeneous CTI reports (using different phrases, verbs, structures). This will result in inaccurate extraction of dependencies.
\par
To address the above challenges, We design a customized NLP pipeline to correctly identify entities, extract dependencies and construct ASGs as follows:% Briefly, the main objectives of this pipeline are two: correct extraction of triples and composition of triples. The specific processes are as follows:
\par
Step 1: Verbosity filtering. We use sentence tokenizer to divide the text into sentences, and then employ the self-trained verbosity filtering model to filter the attack irrelevant sentences. We then retrained the BERT model \cite{devlin2018bert} to filter the irrelevant texts. The training process, parameters and the performance of model are in Section~\ref{subsec:asg_extraction} and Section~\ref{sec:evaluation}. (C1)
\par
Step 2: Attack-related entities identification and protection. Inspired by \cite{huang2015bidirectional, radford2018improving, lample2016neural}, we construct a new BERT-BiLSTM-CRF model for attack-related entity type recognition (first column 'Entity' in Table~\ref{table:system_entities}). The BERT model \cite{devlin2018bert} can enhance the semantic representation of sentences, while the bidirectional LSTM network (BiLSTM) \cite{graves2013speech} can automatically learns the intrinsic connections in the sentence context, and the Conditional Random Field (CRF) \cite{lafferty2001conditional} can restrict the syntax and ensure the plausibility of the predicted labels in terms of order. Then, we combine regular expressions \cite{satvat2021extractor} to subdivide entity attributes (second column 'Symbols' in Table~\ref{table:system_entities}) and replace entities with them to protect the IOC from being mishandled by the NLP modules. (C2)
% The above replacement can improve the follow-up performance of NLP modules applied to traditional text, as well as protect the IOC information from being mishandled by NLP modules. (C2)
\par
Step 3: %Anaphora resolution. In this paper, anaphora refers to explicit anaphora (i.e. explicit reference common in natural language texts). For example, use it/them etc. to refer to previously mentioned entities. 
Anaphora resolution. We use a popular pronoun resolution model, NeuralCoref \cite{neuralcoref}, to map and replace pronouns (e.g., it/that) with previously mentioned entities they refer to. (C2)

\par 
Step 4: Dependency Extraction. After reading a large number of CTI reports, we summarized the types of events necessary to construct an ASG, as shown in Table~\ref{table:system_event}. In order to extract dependencies correctly from CTI reports, the key insight we rely on is that the types of subject-object pair for different events are distinctive, e.g., Write: P --{}\textgreater F, Receive: S --{}\textgreater P, with the exception of "Read" and "Execute" events (both of which are F --{}\textgreater P). In other words, \textbf{after identifying the types of attack-related entity pairs and the interaction direction, we can determine dependencies directly.} Thus, for all attack-related entities in each sentence, we combine them in pairs (discarding impossible entity pairs, e.g., F --{}\textgreater F), and identify the subject and object in these entity pairs (determine the interaction direction) by using part-of-speech (POS) annotation \cite{brill1992simple} and dependency tree \cite{culotta2004dependency}. In addition, for events ("Read" and "Execute") where the subject is of type F and the object is of type P, we vectorize them \cite{mikolov2013word2vec} to calculate the Euclidean distance separately, and then choose the smaller one as the dependency type. By adopting the above mechanism, we construct multiple triples \textless sub, verb, obj\textgreater. (C3)

\par
% Step 8: Co-reference Resolution. Even if the subjects/objects in different triples obtained in the previous step point to the same entity in real world. There are differences in names at the character level due to the description means or model representation. Such reference exists out of context (e.g. Host in CTI reports), so co-reference resolution is needed. For each entity, we iteratively calculate the similarity between itself and the subject or object of other triples. Rename the entity when the similarity exceeds a threshold. The similarity between two entities N and M is calculated as follows: we consider the differences in naming $N_{name}$ between entities, the index $N_{index}$ of the triple where the entity is located, and the type $N_{type}$ between entities, where $W_{d}$ and $W_{t}$ are preset weights. (C2)
Step 5: Co-reference Resolution. Even if entities in different triples are the same, there will be differences in names due to human descriptions. For each entity N, we iteratively calculate the similarity score between N and the entities from other triples. We rename N when the similarity score exceeds the threshold. We consider the differences in naming $N_{name}$ between entities, the index $N_{index}$ of the triple where the entity is located, and the type $N_{type}$ between entities, where $W_{d}$ and $W_{t}$ are preset weights. The similarity score between entity N and M is calculated as follows: (C2)
{\setlength\abovedisplayskip{0.1cm}
\setlength\belowdisplayskip{0.1cm}
\begin{equation}
\resizebox{0.9\hsize}{!}{
$Sim(N, M) = \text{sim}(N_{\text{name}}, M_{\text{name}}) - \frac{|N_{\text{index}} - M_{\text{index}}|} {W_d} - \frac{|N_{\text{type}} - M_{\text{type}}|} {W_t}$
}
\end{equation}

\begin{figure*}[t]
% \caption{Graph Forecast Model Architecture. The current forecast results can be used as input for the next round of prediction for serialized graph generation.}
% \label{fig:graph forecast model archi}
\centering
\includegraphics[scale=0.18]{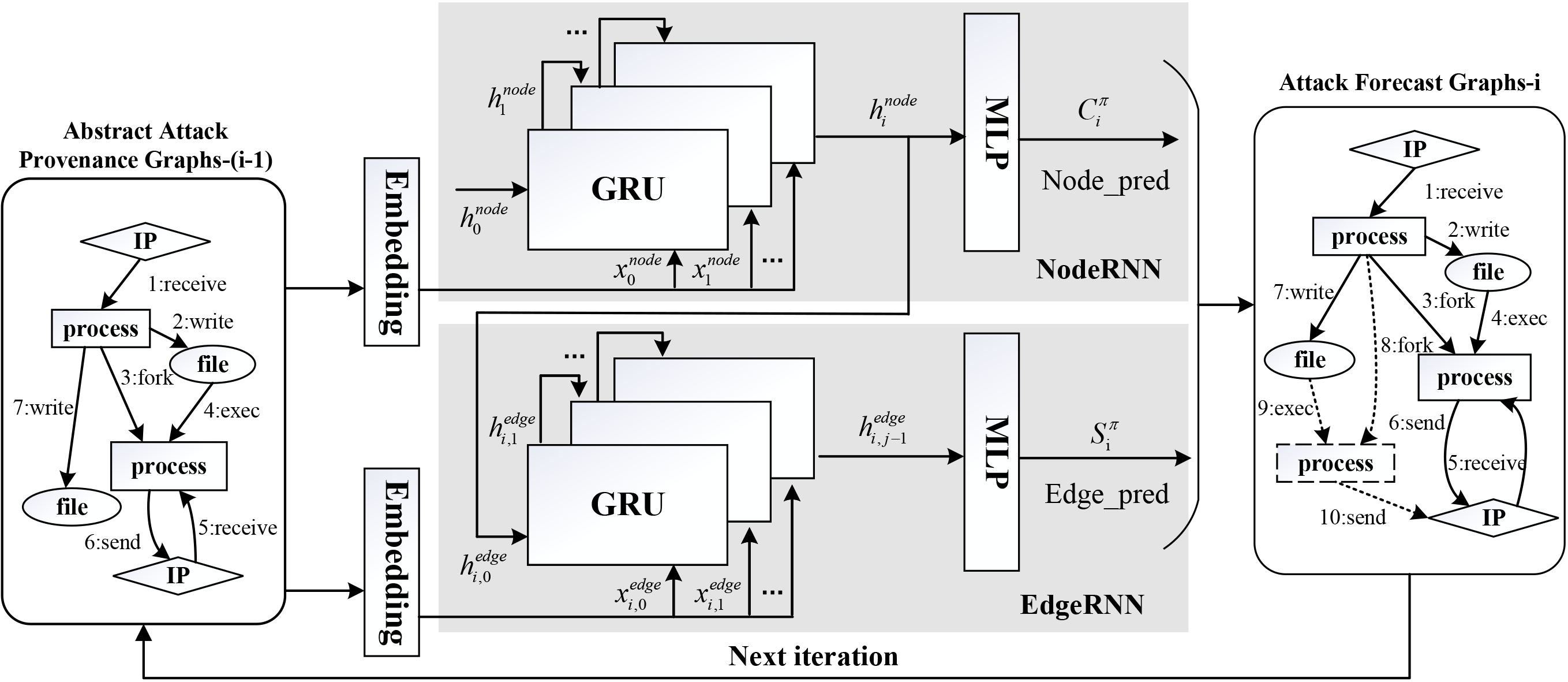}
\setlength{\abovecaptionskip}{0.3cm}
\caption{Graph Forecast Model Architecture. The current forecast results can be used as input for the next round of prediction for serialized graph generation.}
\label{fig:graph_forecast_model_archi}
\end{figure*}

\subsection{AFG Forecast}
\label{sec:4_2}
\par 
In this section, we primarily focus on how to use APGs to generate AFGs.
\par
% \textbf{Goal and Challenge}. The goal of this module is to learn the distribution $p_{model}(G)$ of extracted ASGs $\mathbb{G} = \{G_{1},...,G_{s}\}$ based on the observation over them, where each $G_{i}$ has different interactions. We use the learned distribution to generate graph structure with a certain number of steps for arbitrarily structured graph inputs, thus enabling the generation of AFG. Inspired by You et al.’s work on GraphRNN \cite{you2018graphrnn}, we generate AFGs serially with a deep auto-regressive model. However, challenges still exist. First, GraphRNN only considers homogeneous graphs, but not heterogeneous graphs in which node attributes and edge attributes represent different interactions between different types of entities (we want to generate AFGs that are heterogeneous and can be mapped to the low-level system). For heterogeneous graphs, the node and edge attributes are crucial for the graph structure generation (e.g. there is no direct interaction between two file entities in the system). Second, the inability to converge due to the sparsity of the adjacency matrix during training is not addressed in GraphRNN (e.g. only 200 parts of a 40x40 adjacency matrix have values). For these challenges, we extend the usage of GraphRNN model to heterogeneous graphs and combine the average number of nodes of technique templates (see Appendix~\ref{sec:appendix_imple} for details) to shrink the matrix for sparsity mitigation.

\textbf{Goal and Challenge}. The goal of this module is to learn the distribution $p_{model}(G)$ of extracted ASGs $\mathbb{G} = \{G_{1},...,G_{s}\}$, and use the learned distribution to generate graph structure with a certain number of steps for arbitrarily structured graph inputs, thus enabling the generation of AFG. Inspired by You et al.’s work on GraphRNN \cite{you2018graphrnn}, we generate AFGs serially with a deep auto-regressive model. However, we meet several challenges. First, GraphRNN only considers homogeneous graphs, but for heterogeneous graphs, the node and edge attributes are crucial for the graph structure generation (e.g., there is no direct interaction between two file entities in the system). Second, it is difficult for GraphRNN to solve the non-convergence problem caused by the sparsity of the adjacency matrix during training (e.g., there are only 200 available values in a 40x40 adjacency matrix). For these challenges, we extend the usage of GraphRNN model to heterogeneous graphs and combine the average number of nodes in technique templates (see Section~\ref{sec:implement} for details) to shrink the matrix for sparsity mitigation.
\par
\textbf{Definition and Methodology}.
The attack graphs in this paper are specifically defined as G = (V, E), where the set of nodes is $ V = (v_{1},... ,v_{n}) $,  and the set of edges between nodes is $ E = ({v_{i}, v_{j} |v_{i}, v_{j} \in V}) $. According to certain order of nodes $\pi$, the graph can be represented in the form of adjacency matrix $ A^{\pi} \in \{{S_{i,j}^{\pi} }\} ^{n\times n} $. To illustrate, $\pi$ is a substitution function on $V$ (i.e., $(\pi(v_{1}), ... , \pi(v_{n}) )$), the adjacency element ${S_{i,j}^{\pi} }\in {\{0,1,2,... ,P\}} $ denotes the interaction type between system entities, and the adjacency vector $ S_{i}^{\pi} = (S_{1,i}^{\pi},... ,S_{i-1,i}^{\pi} )^{T} $ denotes the interaction between node $\pi(v_{i})$ and the previous nodes in $\pi$ order. Moreover, the node type ${C_{i}^{\pi} }\in {\{0,1,2,... ,K\}} $ indicates that the node belongs to a certain entity attribute, and the list of all nodes is $C^{\pi} = (C_{1}^{\pi},... ,C_{n}^{\pi})$.
\par
In order to generate graph structures serially, it is crucial to represent graphs with different node orderings as sequences and then build auto-regressive generative models on these sequences. Therefore, we first define a mapping function $f_{S}$ to map graphs to sequences:
{\setlength\abovedisplayskip{0.1cm}
\setlength\belowdisplayskip{0.1cm}
\begin{equation}\label{p(spi)}
p(S^{\pi},C^{\pi})= f_{S}(G, \pi) = ((S_{1}^{\pi},C_{1}^{\pi}), ...,  (S_{n}^{\pi},C_{n}^{\pi}))
\end{equation}}
Where $G \in \mathbb{G}$, G contains n nodes and is represented by the order of nodes $\pi$. At this point, the graph G can be represented by a fixed sequence $S^{\pi}$ and $C^{\pi}$, and the distribution $p(G)$ that we want to learn on the graph can be represented as $p(S^{\pi},C^{\pi})$. Because of the serial property of $\pi$, we can split it into the following likelihood function:

\makeatletter
\renewcommand{\maketag@@@}[1]{\hbox{\m@th\normalsize\normalfont#1}}%
\makeatother

{\setlength\abovedisplayskip{0.1cm}
\setlength\belowdisplayskip{0.1cm}
\begin{small}
\begin{equation}\label{p(Spi,Cpi)}
p(S^{\pi},C^{\pi})= \prod_{i=1}^{n+1} p(C_{i}^{\pi}\mid S_{<i}^{\pi}, C_{<i}^{\pi})p(S_{i}^{\pi} \mid C_{i}^{\pi},S_{<i}^{\pi}, C_{<i}^{\pi} )
\end{equation}
\end{small}
}

First, the graph information consisting of the preceding i-1 nodes is utilized to predict the attribute of the new node. Subsequently, the information from the original graph with i-1 nodes and the predicted node are merged together to forecast the dependence between the predicted node and its predecessor. Note that for the termination node n+1, $p(C_{n+1}^{\pi}\mid S_{<n+1}^{\pi}, C_{<n+1}^{\pi})\equiv 1$.
\par
We refer to the function that predicts the type of a new node as NodeRNN and the function that predicts the dependency between a new node and its predecessor as EdgeRNN. In our model, once the information of the previous i-1 nodes under the node order $\pi$ is determined (i.e., the subgraph/APG is determined), NodeRNN is able to determine the next node i's node type $C_{i}^{\pi}$. And based on the graph information composed of the previous i-1 nodes and the output of NodeRNN $h_{i}^{node}$, the model EdgeRNN is able to generate the dependency relationship between node i and its predecessor node (i.e., the adjacency vector $S_i^{\pi}$). The structure of the whole graph forecast model is shown in Figure~\ref{fig:graph_forecast_model_archi},  the specific parameter settings are shown in Section~\ref{subsec:afg_forecast}, and the specific formulas are as follows:
{\setlength\abovedisplayskip{0.1cm}
\setlength\belowdisplayskip{0.1cm}
\begin{equation}
\resizebox{0.9\hsize}{!}{
$\begin{aligned}
&\text{input}_{i-1} = \text{concat}\left[\text{emb}(S_{i-1}^{\pi}), \text{emb}(C_{i-1}^{\pi})\right] \\
&h_{i}^{\text{node}}, C_{i}^{\pi} = \text{NodeRNN}(h_{i-1}^{\text{node}}, \text{input}_{i-1}), \quad h_{0}^{\text{node}} = 0 \\
% &\psi_{i} = \text{NodeMLP}(h_{i}^{\text{node}}) \\
&h_{i,j}^{\text{edge}}, S_i^{\pi} = \text{EdgeRNN}(h_{i, j-1}^{\text{edge}}, \text{emb}(S_{i,j-1}^{\pi})), \quad h_{i,0}^{\text{edge}} = h_{i}^{\text{node}} \\
% &\phi_{i, j} = \text{EdgeMLP}(h_{i,j}^{\text{edge}})
\end{aligned}$
}
\end{equation}
}
Where concat refers to the stitching between tensors. Since the attack graph is chronological (all system events occur in a backward and forward order in low-level system logs), the node order $\pi$ of the attack graph is determined. 
% See Section~\ref{sec:6.5} for evaluation experiments.
% \begin{figure*}[]
% \caption{Graph Forecast Model Architecture. The current forecast results can be used as input for the next round of prediction for serialized graph generation.}
% \label{fig:graph forecast model archi}
% \centering
% \includegraphics[scale=0.18]{Fig/graph3.png}
% \end{figure*}

\par
% Therefore, we use the model to fully learn the node attributes, edge attributes, temporality, and inter-node dependencies of the ASG, ensuring the validity of the learned distribution. 
% It should be noted that the ASG is used in the model training process, and the APG is used in the real-world combat, both of them are heterogeneous graph structures that can be mapped to the low-level system logs.

\subsection{AFG Interpretation}
\label{sec:4_3}
\par 
% In this section, we introduce how to interpret the AFG generated from the previous pipeline. The technique-level interpretation of the AFG has the following advantages: First, the AFG  generated by the APG which is caused by the possible false positives POI of EDR would be uninterpretable. By false positives reduction, the labor cost can be decreased; Second, after reducing false positives, providing security practitioners with direct attack prediction by interpreting AFG from the technique-level can help in the identification of defense hardening points.
\par
% Concretely, we first provide a generic description of the attack technique and design the ATGs. Then, we use ATGs to interpret AFGs at the attack technique-level (i.e. determine whether the ATG exists in the AFG), but both ATGs and AFGs are abstract (as mentioned in the previous Section~\ref{subsec:3.1}). Inspired by the graph alignment algorithm in poirot \cite{milajerdi2019poirot}, we redesigned and improved the algorithm by proposing \textit{graph alignment plus algorithm}, which extends the algorithm to the abstract level to compute the alignment score in terms of node attributes, edge attributes, and graph structure, while considers the \textbf{Multi-Hop Equivalent Semantics} problem, which is common in attacks. Finally, we present how to use the algorithm results to interpret the AFGs in the form of attack techniques.
In this section, we first introduce how to construct ATGs. Then, we use ATGs to interpret AFGs at the attack technique-level (i.e., determine whether the ATG exists in the AFG). Inspired by the graph alignment algorithm in Poirot \cite{milajerdi2019poirot}, we redesigned and improved the algorithm by proposing \textit{graph alignment plus algorithm}, which extends the algorithm to the abstract level to compute the alignment score in terms of node attributes, edge attributes, and graph structure, while considers the \textbf{Multi-Hop Equivalent Semantics (MHES)} problem, which is common in attacks.
\par
% \textbf{ATG construction:} Atomic red team \cite{atomic-red-team} is a test library proposed by Red Canary that can be mapped to MITRE ATT\&CK, which contains natural language descriptions for different attack techniques and multiple means to implement theses techniques, and it is also called atomic test (currently 265 in total). We manually draw the ATG for each atomic test by reviewing the descriptions of all the atomic red team techniques and the command lines of the test cases, as shown in Appendix~\ref{sec:appendix_atgexam}. The ATGs reflect the form of low-level system log of the attack techniques with multiple nodes and edges, where nodes represent system entities and edges indicate the events. We have made these technique templates open source on website \footnote{open-source website}, where the relevant statistical data are shown later in Appendix~\ref{sec:appendix_imple}.
\textbf{ATG construction:} Atomic red team \cite{atomic-red-team} is a test library proposed by Red Canary that can be mapped to MITRE ATT\&CK. It contains descriptions and implementations of different attack techniques (atomic tests, currently 265 in total). By reviewing these, we manually construct the ATG for each atomic test, as shown in Appendix~\ref{sec:appendix_atgexam}. ATGs reflect the form of low-level system log of the attack techniques with multiple nodes and edges, where nodes represent entities and edges indicate events. And the relevant statistical data about ATGs are shown later in Section~\ref{subsec:afg_interpretation}.
% We have made these technique templates open source on website \footnote{open-source website}, where the relevant statistical data are shown later in Appendix~\ref{sec:appendix_imple}.

\begin{table*}[t]
\caption{Suspicious Semantic Delivery Rules.}
\label{table:suspecious_semantice_rule}
\centering
\scalebox{1.2}{
\begin{tabular}{|c|c|c|c|}
\hline
\textbf{Event Type} &
  \textbf{Subject} &
  \textbf{Object} &
  \textbf{Description} \\ \hline
Read &
  Process &
  File &
  \begin{tabular}[c]{@{}c@{}}The process reads a file containing suspicious semantics \\ and the process is considered to be associated with the attack.\end{tabular} \\ \hline
Write &
  Process &
  File &
  \begin{tabular}[c]{@{}c@{}}A process associated with the attack writes a file, \\ and the file being written contains suspicious semantics.\end{tabular} \\ \hline
Load &
  Process &
  File &
  \begin{tabular}[c]{@{}c@{}}The process that loaded files containing suspicious \\ semantics are considered to be associated with an attack.\end{tabular} \\ \hline
Receive &
  Process &
  Socket &
  \begin{tabular}[c]{@{}c@{}}The process is considered to be associated with \\ the attack after accepting data from the suspicious network.\end{tabular} \\ \hline
Send &
  Process &
  Socket &
  The process associated with the attack sends data outward. \\ \hline
Fork/Clone &
  Process &
  Process &
  \begin{tabular}[c]{@{}c@{}}The process spawns a new process through the Fork/Clone \\ operation and passes suspicious semantics to the new process.\end{tabular} \\ \hline
\end{tabular}}
\end{table*}

\par
\textbf{Graph alignment plus algorithm:} We mainly calculate the alignment score $\Gamma (G_q : : G_p)$ by applying the graph alignment plus algorithm to quantitatively determine the presence of the ATG ($G_q$) in the AFG ($G_p$). And the implementation of the algorithm can be summarized in three steps: 
1) Find candidate nodes. Find all corresponding candidate nodes k in $G_p$ for all nodes $i$ in $G_q$, respectively, and we call the set of candidate nodes corresponding to $i$ as $\Gamma_C(i)$. 
2) Fix candidate nodes. Iterate through the node scores of all candidate nodes of node $i$, select the node $m$ with the highest score and greater than the threshold value as the fixed node of node $i$ ($m \in \Gamma_C(i)$ ), we note the fixed node corresponding to $i$ as $\Gamma_F(i) $.
3) Compute the graph alignment score. After determining the fixed nodes of all nodes in $G_q$, we compute the alignment score $\Gamma(G_p, G_q)$ between two graphs.
\par
The method for determining candidate nodes is shown in Equation~\ref{equation:node_candidate}, and if $Candi (i:k)$ is 1 then $k$ is added to $\Gamma_C(i)$, where $i_{degree}$ denotes the degree of node $i$.
The node score of the candidate node is calculated as shown in Equation~\ref{equation:node_score}, where $(i \Longrightarrow j) $ denotes the path starting with $i$ and ending with $j$, both $i$ and $k$ are fixed as inputs.
The path score is calculated as shown in Equation~\ref{equation:path_socre}, where $i',j'$ is the node in path $i \Longrightarrow j$, $i' \to j'$ is an edge formed by $i',j'$, and $(i' \to j')$ is an edge in path $i \Longrightarrow j$.
The edge score is calculated as shown in Equation~\ref{equation:edge_socre}, where judgment conditions of MHES will be introduced later.
Finally, the graph alignment score $\Gamma(G_q, G_p)$ is computed as shown in Equation \ref{equation:graph_align}, which represents a quantitative determination of the presence of $G_q$ in $G_p$.
% \makeatletter
% \renewcommand{\maketag@@@}[1]{\hbox{\m@th\normalsize\normalfont#1}}%
% \makeatother
{\setlength\abovedisplayskip{0.1cm}
\setlength\belowdisplayskip{0.1cm}
{
\begin{small}
\begin{align}\label{equation:node_candidate}
&Candi (i:k) = \begin{cases}
    1, & i_{type} =  k_{type} \wedge i_{degree} <= k_{degree}\\
    0,  & else
\end{cases} 
\end{align}
\end{small}}

\begin{small}
\begin{align}\label{equation:node_score}
&NodeScore(i, k) = \sum_{j \in G_q} \frac{PathScore(i, j: k)}{(i \Longrightarrow   j)} 
\end{align}
\end{small}

\begin{small}
\begin{align}\label{equation:path_socre}
&PathScore(i, j:k) = \sum_{\substack{i' ,j' \in i \Longrightarrow j \\ i' \to j' \in  i \Longrightarrow j} } \frac{EdgeScore(i', j': k)}{(i' \to   j')} 
\end{align}
\end{small}

\begin{small}
\begin{equation}
\begin{aligned}\label{equation:edge_socre}
&EdgeScore(i', j':k) = \sum_{\substack{l \in \Gamma_c^{j} \\ k,l \in G_p} } \frac{MHES(i',j':k,l)}{(k \Longrightarrow l)}  \\
&MHES (i', j':k, l) = \begin{cases} 1, 
    & i' \to j'  \ and \ k \Longrightarrow j \ with \\ 
    & \ Equivalent \ Semantics\\ 0,  
    & else
\end{cases} 
\end{aligned}
\end{equation}
\end{small}

\begin{small}
\begin{align}\label{equation:graph_align}
&\Gamma(G_p : G_q) = \sum_{i,j \in G_q} \frac{(\Gamma_F(i) \Longrightarrow \Gamma_F(j) )}{(i \Longrightarrow j)}  
\end{align}
\end{small}
}
\par
% Regarding \textbf{MHES}, we believe that during the attack, the interaction between the suspicious process and other entities would make its own suspicious information flow and control flow be passed according to certain systematic rules, resulting in suspicious semantics of other entities (i.e., semantic transfer in forensic analysis \cite{xiong2020conan}). ATG is applied by EFI to interpret AFG at the technique-level, but ATG is extremely concise due to its atomistic characteristics, and if it is directly applied to graph alignment detection for interpretation, there would be high false negatives (i.e. the semantics are the same but cannot be identified), which would lead the attacker to "go the long way" and easily evade.
% As Figure~\ref{fig:Multi-hop semantic equivalence diagram} shown, in the right diagram, the suspicious semantics of control flow and information flow of process P1 has passed to P3 with the interaction between entities after multiple operations, then the semantics of this multi-hop path should be equivalent to the single-hop in the left diagram, which represents the proposed concept of Multi-Hop Equivalent Semantics.
We find that ATG is usually streamlined due to its atomic characteristics. If it is directly applied to graph alignment for interpretation, there will be a high number of false negatives. Also, it will be easily avoided by ``take a long way" attackers. Fortunately, suspicious semantics of entities will be passed through control flow and information flow \cite{xiong2020conan,zhu2021general}. As shown in Figure~\ref{fig:Multi-hop_semantic_equivalence_diagram}, the suspicious semantics of process P1 in Figure~\ref{fig:Multi-hop_semantic_equivalence_diagram_b} will pass to P3 with control flow (t = 1) and information flow (t = 2, 3), then the semantics of this multi-hop path should be equivalent to the single-hop in Figure~\ref{fig:Multi-hop_semantic_equivalence_diagram_a}, which represents the proposed concept of MHES.
\par
To determine whether there is semantic equivalence between a single-hop edge $i \to j$ and a multi-hop path $k \longrightarrow l$, we summarize the suspicious
semantic delivery rules through observing the delivery phenomenon of semantic caused by system entities when passing information flows and control flows, as shown Table~\ref{table:suspecious_semantice_rule}. We define single-hop edges as $i \to j = (<v_i, v_j,e_v>)$, n-hop paths as
$k \longrightarrow l = (<v_1,v_2,e_1>, <v_2,v_3,e_2>,... ,<v_{n-1},v_n,e_n>)$, expressing the edge information in terms of triples $<subject, object, event type>$, with the table of suspicious semantic delivery rules defined as the set $U$. Then, if
{\setlength\abovedisplayskip{0.2cm}
\setlength\belowdisplayskip{0.2cm}
\begin{align}
\begin{split}
&\exists i.type == v_1.type \wedge j.type == v_n.type \\ &\wedge e_v == e_n \wedge <v_{t-1},v_t,e_t> \in U,\forall t \in[1,n)
\end{split}
\end{align}}
We consider the edge $i \to j$ and the multi-hop path $k \longrightarrow l$ are semantic equivalent, and the MHES judgment function in Equation~\ref{equation:edge_socre} returns 1.

\begin{figure}[ht]
% \caption{An example of Multi-Hop Equivalent Semantics.}
%\label{fig:Multi-hop semantic equivalence diagram}
\centering
\subfigure[One-hop Edge]
{
    % \begin{minipage}[One-hop Edge]{1.0\linewidth}
        \includegraphics[scale=0.5]{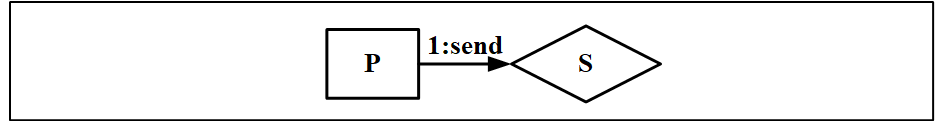}
    % \end{minipage}
    \label{fig:Multi-hop_semantic_equivalence_diagram_a}
}

\subfigure[Multi-hop Edge]
{
    % \begin{minipage}[Multi-hop Edge]{1.0\linewidth}
        \includegraphics[scale=0.5]{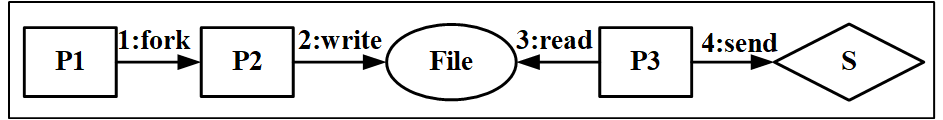}
    % \end{minipage}
    \label{fig:Multi-hop_semantic_equivalence_diagram_b}
}
\setlength{\abovecaptionskip}{-0.1cm}
\caption{An example of Multi-Hop Equivalent Semantics.}
\label{fig:Multi-hop_semantic_equivalence_diagram}
\vspace{-0.2cm}
\end{figure}

% \begin{table*}[]
% \caption{Suspicious Semantic Delivery Rules Table}
% \label{table:suspecious_semantice_rule}
% \centering
% \scalebox{1.2}{
% \begin{tabular}{|c|c|c|c|}
% \hline
% \textbf{Event Type} &
%   \textbf{Subject} &
%   \textbf{Object} &
%   \textbf{Description} \\ \hline
% Read &
%   Process &
%   File &
%   \begin{tabular}[c]{@{}c@{}}The process reads a file containing suspicious semantics \\ and the process is considered to be associated with the attack.\end{tabular} \\ \hline
% Write &
%   Process &
%   File &
%   \begin{tabular}[c]{@{}c@{}}A process associated with the attack writes a file, \\ and the file being written contains suspicious semantics.\end{tabular} \\ \hline
% Load &
%   Process &
%   File &
%   \begin{tabular}[c]{@{}c@{}}The process that loaded files containing suspicious \\ semantics are considered to be associated with an attack.\end{tabular} \\ \hline
% Receive &
%   Process &
%   Socket &
%   \begin{tabular}[c]{@{}c@{}}The process is considered to be associated with \\ the attack after accepting data from the suspicious network.\end{tabular} \\ \hline
% Send &
%   Process &
%   Socket &
%   The process associated with the attack sends data outward. \\ \hline
% Fork/Clone &
%   Process &
%   Process &
%   \begin{tabular}[c]{@{}c@{}}The process spawns a new process through the Fork/Clone \\ operation and passes suspicious semantics to the new process.\end{tabular} \\ \hline
% \end{tabular}}
% \end{table*}

\par
After calculating the alignment scores of all the technique templates, we compare them with the predefined threshold $T$ and consider that the technique templates $G_t$ exceeding the threshold exist in the AFG $G_f$, achieving the interpretation of AFG at the technique-level.
% \par
% The above is the general framework of EFI, where the implementation details and specific tools are shown in the Appendix~\ref{sec:appendix_imple}.

\subsection{Strategy Dispatch}
\label{sec:4_4}
% In the real scenario, EFI, as a third-party tool to the existing EDR, needs to use the APG provided by the EDR as input to the graph generation model (as shown in Section~\ref{sec:3.3}). Therefore, we will not describe details of the APG investigation; all APGs mentioned in this paper are provided by existing EDRs, and we simulate this situation by reproducing other work in Section~\ref{sec:6.5.3}. Based on the generation of AFG, EFI will dispatch strategy to EDR, such as predicting \textless P, read, F0\textgreater to prohibit the process from reading sensitive files (combined with whitelisting), and predicting \textless P, send, S\textgreater to prohibit the process from communicating with sockets.  In addition, these strategies can be combined with heuristic decay, i.e., lifting the restriction after a certain period of time.

\textbf{Graph Investigation}. In practical usage, deviations in the prediction results may lead to the blocking of normal behavior. To address this problem, EFI will analyze AFGs by adopting existing attack detection/investigation systems. Here, we can utilize the TTP-based context analysis model proposed in APTSHIELD \cite{zhu2023aptshield} to examine the integrity and rationality of the AFG (i.e., technique-level analysis), and only AFGs that align with the complete attack flow will be blocked at forecast points. This approach maximizes the value of traditional APT detection systems and transforms their response capabilities into enhanced predictive reinforcement capabilities. Furthermore, we can integrate multiple real-time models (e.g., TTP-based \cite{zhu2023aptshield}, anomaly-based \cite{han2020unicorn}, classification-based \cite{chen2022apt}) to inspect whether AFG constitutes a comprehensive attack, thereby minimizing FPs of forecast results to the greatest extent possible.

\textbf{Automated Reinforcement}. Based on the generation of AFG, EFI will dispatch strategy to EDR, such as predicting \textless P, read, F0\textgreater to prohibit the process from reading sensitive files (combined with whitelisting), and predicting \textless P, send, S\textgreater to prohibit the process from communicating with sockets.  In addition, these strategies can be combined with heuristic decay, i.e., lifting the restriction after a certain period of time.

\section{Implementation}\label{sec:implement}

In this section, we will briefly describe the implementation details and the dependent tools in our system. We deployed EFI on a Windows system with Intel (R) Core (TM) i9-10900K CPU @ 3.70GHz and 64GB memory.
\subsection{ASG Extraction}
\label{subsec:asg_extraction}
\par
We used NLTK~\cite{nltk} for sentence splitting, spaCy~\cite{spacy} for POS annotation and dependency tree construction, NeuralCoref~\cite{neuralcoref} for resolving anaphora. We subsequently optimized models to filter the irrelevant texts and identify the attack-related entities. Specifically, we manually labeled 8,000 sentences and tagged 10,451 entities from blogs of 5 security vendors(bitdefender~\cite{bitdefender},
Microsoft~\cite{microsoft_blog},
symantec~\cite{symantec_blog},
talosintelligence~\cite{talosintelligence_blog}, and virustotal~\cite{VirustotalBlog}). For the methodology of sentence labeling,  we followed the purpose of text classification to label them under two classes of attack-relevant (label=1) and attack-irrelevant (label=0). For the methodology of entity tagging, we used the BIO scheme~\cite{ramshaw1999text}, where the B-prefix indicates the beginning of the tag, and the I-prefix indicates the inside of the tag. An outside (O) tag is a token indicating that it is not in a predefined category. More detailed statistics on the dataset are presented in Table~\ref{table:statistic_dataset_sentence_and_entity}.

\begin{table}[t]
\centering
\caption{The detailed statistics about the labeled sentence dataset and the tagged entity dataset.}
\label{table:statistic_dataset_sentence_and_entity}
\begin{tabular}{|c|c|c|c|}
\hline
\textbf{Data Type}                 & \textbf{Data Label} & \textbf{Count} & \textbf{Ratio(\%)} \\ \hline
\multirow{3}{*}{\textbf{Sentence}} & Attack-Relevant     & 4000           & 50                \\
                                   & Attack-Irrelevant   & 4000           & 50                \\ \cline{2-4} 
                                   & Total               & 8000           & 100               \\ \hline
\multirow{4}{*}{\textbf{Entity}}   & File                & 3159           & 30.2              \\
                                   & Process             & 4845           & 46.4              \\
                                   & Socket              & 2447           & 23.4              \\ \cline{2-4} 
                                   & Total               & 10451          & 100               \\ \hline
\end{tabular}
\vspace{-0.2cm}
\end{table}

\par
We then used the sentences to train the BERT model for redundancy filtering. Besides, we also used the entities to train the BERT-BiLSTM-CRF model for attack-related entity identification. The detailed evaluation is shown in Section~\ref{sec:6_3} and Section~\ref{sec:6_4}. 
Finally, we used graphviz~\cite{graphviz} to visualize the triples and networkX~\cite{networkx} to save the adjacency matrix with more than 5 nodes (we consider that a basic information collection and leakage involves at least 5 nodes), and finally collected 1,429 attack scene graphs, containing a total of 19,212 entities and 27,586 inter-entity dependencies.

\subsection{AFG Forecast}
\label{subsec:afg_forecast}
\par
We constructed two sub-models, namely NodeRNN and EdgeRNN. In NodeRNN, the adjacency matrix and attributes of predecessor nodes served as the input, which underwent embedding layers and expanded into 64 and 256 dimensional outputs, respectively. Then outputs were concatenated and fed into the GRU network with 4 hidden layers and 128 dimensions. The resulting output $h_{output}^{node}$ was deployed as input for the EdgeRNN and node attribute prediction. Subsequently, EdgeRNN accepted $h_{output}^{node}$ as input and expanded it into 32 dimensions via the embedding layer before inputting it into the GRU network with 4 hidden layers and 64 dimensions. Finally, linear layers transformed the output $h_{output}^{edge}$ to obtain predictions of the edge relationship between node \textit{i} and its predecessor nodes. In addition, the experiments were conducted with a batch size of 16 and for 50 epochs, while the dataset (ASGs) was divided into 1,143 for training, and 286 for testing. The detailed evaluation are shown in Section~\ref{sec:6_5}.

\subsection{AFG Interpretation}
\label{subsec:afg_interpretation}
\par
We used the description of the techniques and actual code examples by the atomic red team \cite{atomic-red-team} to build ATGs manually. We ended up with a total of 256 ATGs for each of the available atomic techniques. In total, 882 process nodes, 361 file nodes, 52 registry nodes, 37 socket nodes and 1,192 inter-node dependencies were involved, covering 12 tactics, 123 techniques and 256 sub-techniques, with an average of 5.2 nodes and 4.7 edges per ATG. Finally, ATGs and \textit{graph alignment plus algorithm} are combined to be used for the interpretation of APGs.

\section{Evaluation}\label{sec:evaluation}
In this section, we evaluate the effectiveness of each component of EFI and answer the following four questions. 
\par
\textbf{RQ1:} How to prove the effectiveness of EFI on ATG 
construction? 
\par
\textbf{RQ2:} How to prove the effectiveness of graph alignment plus algorithm? 
\par
\textbf{RQ3:} How to prove the effectiveness of EFI on ASG extraction? 
\par
\textbf{RQ4:} How to prove the effectiveness of EFI on AFG generation?

% \par
% \begin{itemize}
% \item[$\bullet$] \textbf{R1:} How to prove the effectiveness of attack template graph (ATG) construction?
% \item[$\bullet$] \textbf{R2:} How to prove the effectiveness of graph alignment plus algorithm?
% \item[$\bullet$] \textbf{R3:} How to prove the effectiveness of attack scene graph (ASG) extraction?
% \item[$\bullet$] \textbf{R4:} How to prove the effectiveness of attack forecast graph (AFG) generation?
% \end{itemize}

\subsection{Evaluation Preparation}
\label{sec:6_1}
\par
To evaluate EFI, we collected 3,483 open-source CTI reports, extracted a total of 1,429 ASGs, annotated 10,451 attack-related entities and 8,000 sentences, and constructed 256 ATGs. In addition, we performed manual extraction and technique-level interpretation of ASGs on 10 CTI reports, so as to construct ground truth for the evaluation. The above 10 reports were selected to ensure that they cover different operating systems, various kill chains, including 4 from the Darpa Transparent Computing Dataset \cite{DARPA-TC} and 6 from APT organization reports. The specific report names are shown in Table~\ref{table:asg_NodeEdgeAlignscore_eval}, where TC\_A1-A4 represent hc attack, ccleaner attack, Information gather and exfiltarion, and In-memory attack with firefox in DARPA reports, respectively.
\par
% We explain the order of the problem setup. Firstly, since answering R2 and R4 both require the application of the ATG (i.e., the technique template), we answer R1 to evaluate our technique template, then answering R3 and R4 both require the application of our \textit{graph alignment plus algorithm}, so we will first evaluate the algorithm to answer R2, then evaluate the ASG extraction effectiveness to answer R3, and finally design experiments to evaluate the AFG generation to answer R4.
To evaluate the effectiveness of EFI on ASG extraction, we reproduced two latest works EXTRACTOR~\cite{satvat2021extractor} and AttackG \cite{li2021attackg} to compare with our system. We selected these two systems due to their similar aims to ours, which involve extracting graphs representing attack behaviors from CTI. 
\par
To simulate the role of EDR in real scenarios, as described in Section~\ref{sec:3_3}, we also reproduced DEPIMPACT~\cite{fang2022back} for corelated sub-graph investigation of alert point, and reproduced DEPCOMM~\cite{xu2022depcomm} and CPR~\cite{xu2016cpr} for compressing redundant edges and nodes in the sub-graph. We examined the effectiveness of EFI on AFG generation by passing the pruned ASG into the graph forecast model as APG.
\par
In addition, we simulated the cooperation between EFI and EDR tools in the DARPA Engagement dataset (TC\_A1-A4) collected under real scenarios as a way to demonstrate that EFI eliminates the gap between CTI reports and the system logs for effective forecast.
% and strategy dispatching in syslog-level.

\subsection{RQ1: How to prove the effectiveness of ATG construction?}
\label{sec:6_2}
\par
In our system, ATG is applied to the interpretation of AFG. Also, ATG is the key component when answering R2 and R4. ATGs in EFI are constructed according to a total of 256 panels covering 12 tactics, 123 techniques and 256 sub-techniques of MITRE ATT\&CK.
\par
% we randomly selected 10 seed techniques and ensured that these 10 seed techniques were distributed among different tactics as much as possible, and then implemented these sub-techniques on a real operating system and performed log collection, where for the log collection part we used the SPADE \cite{SPADE} collection system with Auditd \cite{Auditd} as the kernel. 
% For the collected system logs, we first performed irrelevant log filtering, i.e., we only kept the logs of processes and threads related to the implementation of sub-techniques commands, and then looked for entities and dependencies in the technique templates in the logs to see if they had corresponding event records. 
% For the collected system logs, we first filter irrelevant contents to retain entities and events that implemented by atomic tests, and then compare filtered graph with corresponding ATG. 
% After traversing all the logs, we analyzed and found that all the entities and dependencies in our manually drawn technique template could find corresponding event records in the logs, i.e., we proved the effectiveness of the attack template graph. 
To evaluate the effectiveness of the ATG, we randomly selected 50 techniques that were evenly distributed across different tactics, then implemented these techniques (atomic tests) on a real system with SPADE \cite{SPADE} to collect low-level system logs. The experimental results show that all entities and dependencies (triples) can be found in the filtered graphs, which illustrates that the constructed ATGs truly represent technique implementations on low-level system logs.
The specific implementations and  system logs of picked atomic tests can be found in our open-source project.

\subsection{RQ2: How to prove the effectiveness of graph alignment plus algorithm?}
\label{sec:6_3}

\par 
% Before formally evaluating the effectiveness of the graph alignment plus algorithm, we first need to state that the algorithm is only used for AFG interpretation in the normal use of this system (see Section~\ref{sec:4.3} for details). However, since we would use this algorithm in the evaluation as an indicator for the assessment of ASG extraction effectiveness and AFG generation effectiveness to answer R2 and R4, specifically for the quantitative judgment of the alignment scores between two graphs and the qualitative judgment of the technique templates contained in one graph, respectively. Therefore, in this subsection, we will evaluate the effectiveness of the algorithm on the calculation of alignment scores as well as on the interpretation of technique templates.
We need to state that \textit{graph alignment plus algorithm} is not only used for AFG interpretation in the normal use of EFI (see Section~\ref{sec:4_3} for details). But we also use this algorithm to evaluate the effectiveness of ASG extraction (R3) and AFG generation (R4) in experiment. Therefore, we evaluate the effectiveness of the algorithm on the calculation of alignment scores as well as on the technique-level interpretation.
\par
\textbf{Graph alignment score calculation:} The fundamental purpose of this algorithm is to quantify the extent to which the graph structure of $G_q$ is present in $G_p$. While the alignment score exceeds a threshold value we assume that $G_p$ contains $G_q$. We first randomly selected 10 attack graphs from ASGs and set them as $G_p$ one by one, and performed three kinds of operations on each $G_p$: random edge addition and deletion, random node addition and deletion, and node addition and deletion consistent with MHES. For each processed $G_q$, the \textit{graph alignment plus algorithm} was applied separately to calculate the alignment scores between the processed graph $G_q$ and the original graph $G_p$. The average results are shown in Figure~\ref{fig:graphalign_plus}. We can find that \textbf{the impact of adding or deleting an edge on the graph alignment score is much smaller than that of a node}, the reason is that the core of this algorithm lies in computing the NodeScore to fix the candidate nodes, while the addition and deletion of edges only affects a small portion of the pathway. Moreover, \textbf{adding or deleting graph nodes according to MHES has a much smaller impact on the final score than randomly adding or deleting graph nodes}, which is in line with our expectation, since the \textit{graph alignment plus algorithm} avoids false negatives of results by considering equivalent semantics. Finally, we find that \textbf{the addition operation has less impact on the score than that of the deletion operation}, which is due to the fact that deleting a graph node destroys a large number of path structures where that node exists, leading to a rapid decrease in PathSocre. 
% It should be noted that this subsection for the evaluation of the alignment score calculation is applied both to evaluate the effectiveness of ASG extraction in R3 and to support the technique template interpretation function for AFG in R4.
\begin{figure}[ht]
% \caption{Change in graph alignment score with graph structure modification. The vertical axis indicates the alignment score and the horizontal axis indicates the number of modified nodes (edges). The positive number of the horizontal axis indicates an increase, and the negative number indicates a deletion.}
%\label{fig:graphalign+}
\vspace{-0.6cm}
    \centering
    \includegraphics[width=\linewidth]{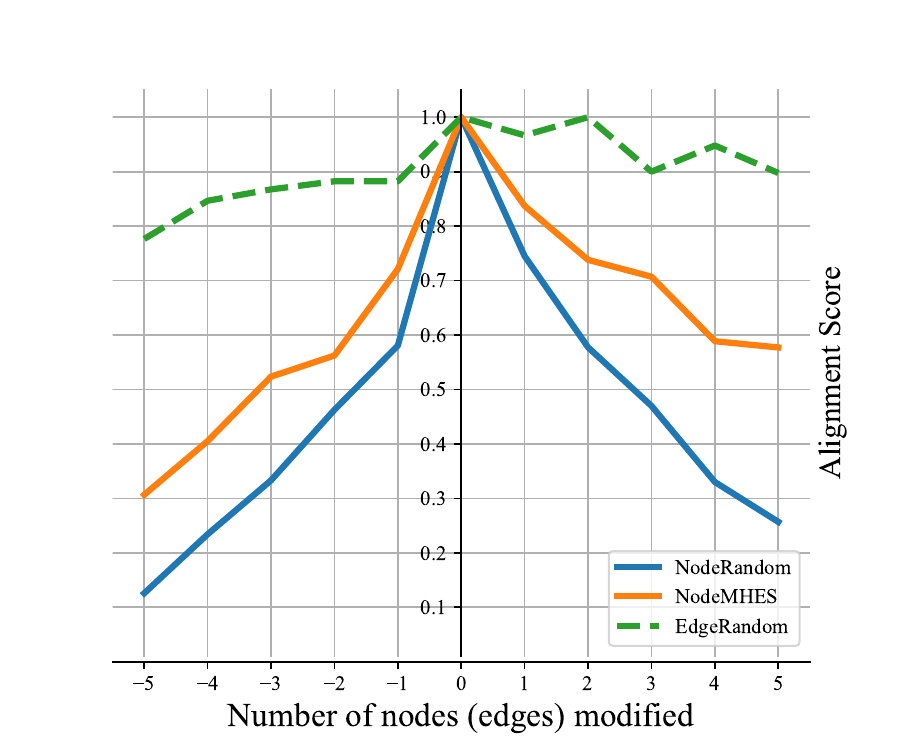}
    \setlength{\abovecaptionskip}{-0.4cm}
    % \caption{Change in graph alignment score with graph structure modification. The vertical axis indicates the alignment score and the horizontal axis indicates the number of modified nodes (edges). The positive number of the horizontal axis indicates an increase, and the negative number indicates a deletion.}
    \caption{Change in graph alignment score with graph structure modification. The positive/negative number of the horizontal axis indicates an addition/deletion.}
    \label{fig:graphalign_plus}
\vspace{-0.3cm}
\end{figure}
\par

\textbf{Technique-level interpretation:} The \textit{graph alignment plus algorithm} is not only applied to the technique interpretation of AFG in real-world scenarios. Since AFG and ASG are the same in graph structure and level, here, we used \textit{graph alignment plus algorithm} to identify the ATGs on the manually annotated ASGs (mentioned in Section~\ref{sec:6_1}) and compared the result with that of latest open-source studies on technique-level interpretation, TTPDrill\footnote{https://github.com/KaiLiu-Leo/TTPDrill-0.5} and AttackG\footnote{https://github.com/li-zhenyuan/Knowledge-enhanced-Attack-Graph}, as shown in Table~\ref{table:asg_NodeEdgeAlignscore_eval}. From Table~\ref{table:asg_NodeEdgeAlignscore_eval}, TTPDrill is the worst in terms of overall performance, while EFI is slightly better than AttackG. After analysis, we find that the average number of extracted techniques are 9.9, 17.8 and 7.3 for AttackG, TTPDrill, and EFI, respectively. TTPDrill extracts a large number of techniques because it generates a technique for each sentence without considering contextual information associations. In addition, we find that the precision of AttackG is higher than recall. The reason is that the graph extracted by AttackG is relatively discrete (e.g., the graph is split into five subgraphs on the CTI report ``Information gather and exfiltration (TC\_A3)" from Darpa TC Dataset, resulting in only one technique being hit). The precision of both TTPDrill and EFI is lower than that of recall (i.e., higher FPs and lower FNs). For TTPDrill, we find that recall is twice as high as precision, the reason is that TTPDrill has an extremely high number of FPs (average 16 FPs per report) and is not interpreted by newer techniques (i.e., new techniques updated by MITRE after 2017). In our design, we prefer to achieve high recall to reduce FNs. Since EFI can dispatch strategies to EDR based on the interpreted results. Even if there are false positives, advance reinforcement will not affect normal operations of the enterprise. However, FNs may cause companies to miss the optimal defence time.

\subsection{RQ3: How to prove the effectiveness of ASG extraction?}
\label{sec:6_4}

\begin{table*}[ht]
\centering
\caption{The performance of the model on redundant text filtering. Columns 2 to 4 indicate the number of original text sentences, filtered text sentences, and attack-related sentences, respectively. Column 5 refers to true positive/false positive/true negative/false negative, and columns 6 to 8 refer to the Precision, Recall and F-1 Score calculated with TP/FP/FN.}
\label{table:bert_redundent_filter}
\begin{tabular}{|c|c|c|c|c|c|c|c|}
\hline
\textbf{\begin{tabular}[c]{@{}c@{}}CTI\\ Reports\end{tabular}} & \textbf{\begin{tabular}[c]{@{}c@{}}Original Text\\ (No. of Sent.)\end{tabular}} & \textbf{\begin{tabular}[c]{@{}c@{}}Filtered Text\\ (No. of Sent.)\end{tabular}} & \textbf{\begin{tabular}[c]{@{}c@{}}Groundtruth Text\\ (No. of Sent.)\end{tabular}} & \textbf{TP/FP/FN} & \textbf{\begin{tabular}[c]{@{}c@{}}Precision\\ (\%)\end{tabular}} & \textbf{\begin{tabular}[c]{@{}c@{}}Recall\\ (\%)\end{tabular}} & \textbf{\begin{tabular}[c]{@{}c@{}}F1-Score\\ (\%)\end{tabular}} \\ \hline
{\color[HTML]{000000} Carbanak}                                & 334                     & 68                       & 49                          & 45/23/3           & 66.18                                                            & 91.84                                                         & 76.93                                                           \\
DeputyDog                                                      & 65                      & 7                        & 5                           & 5/2/0             & 71.43                                                            & 100                                                           & 83.33                                                           \\
DustySky                                                       & 297                     & 37                       & 28                          & 26/11/2           & 70.27                                                            & 92.86                                                         & 80.00                                                           \\
njRAT                                                          & 443                     & 29                       & 24                          & 22/7/2            & 75.86                                                            & 91.67                                                         & 83.02                                                           \\
OceanLotus                                                     & 199                     & 33                       & 29                          & 24/9/5            & 72.73                                                            & 82.76                                                         & 77.42                                                           \\
Uroburos                                                       & 154                     & 19                       & 16                          & 16/3/0            & 84.21                                                            & 100                                                           & 91.43                                                           \\ \hline
\textbf{Average}                                               & 248.5                   & 32.17                    & 25.17                       & -                 & 73.47                                                            & 93.19                                                         & 82.02                                                           \\ \hline
\end{tabular}
\vspace{-0.2cm}
\end{table*}

\par
% In order to answer R3, we first need to state that in this system, the extraction of ASGs is to increase data source, and then use them for training graph forecast model. However, the graph forecast model relies more on obtaining information from graph distribution, and does not care about the specific name of entities. Therefore, our evaluation of ASG extraction focuses on the effectiveness of the module in extracting entity attributes, dependency attributes, and overall graph structure without considering the specific name of entities (i.e., the abstraction mentioned in Section~\ref{sec:4.3}). In addition, as mentioned in Section~\ref{sec:4.1}, there is a large amount of text redundancy in open-source CTI reports. Therefore, this module will be evaluated in these two aspects in this section.
The purpose of extracting ASGs from CTI reports is to obtain a large number of samples that can be used to train the graph forecast model. \textbf{Note that the graph forecast model relies on obtaining information from graph distribution rather than the specific names of entities.} Here, we evaluate the effectiveness of the module in extracting entity attributes, dependency attributes, and overall graph structure, as well as the redundant sentences filtering mentioned in Section~\ref{sec:4_1}.

\par
\textbf{Redundant sentences filtering:} 
Since reports from the Darpa TC \cite{DARPA-TC} (TC\_A1-A4) are already quite concise (only 6 sentences per report on average), we selected 6 APT organization reports for evaluation. We annotated the attack-relevant sentences as the ground truth by referring to the attack graph drawn in Poirot \cite{milajerdi2019poirot}. Then we input 6 CTI reports into our filtering model (Section~\ref{sec:4_1}) to get the filtered text. Finally we computed the evaluation metrics on the filtered text and the attack-related sentences, as shown in Table~\ref{table:bert_redundent_filter}. 
From the table, we can see that the filtering model has a very high performance in terms of recall (i.e., retain attack-related sentences) and a slightly poorer performance in terms of precision (i.e., retain attack-irrelevant sentences). As mentioned in Section~\ref{sec:4_3}, the number of attack-related sentences in a CTI report is extremely smaller than that of attack-unrelated. Therefore, we prefer to reduce false negatives to capture all attack-related sentences in CTI reports. Moreover, after analysis, the poor performance of Carbanak \cite{Carbanak} is due to the fact that it contains much descriptive sentences about the characteristics of malicious processes, e.g., 'Carbanak is a backdoor used by the attackers to compromise the victim's machine once the exploit, either in the spear phishing email or exploit kit, successfully executes its payload.'

\par
\textbf{ASG extraction:} 
% In order to evaluate the effectiveness of the ASG extracted by this module, we first performed manual extraction of ASGs for 10 reports as ground truth. In the drawing process, we both consider the descriptions in the report \cite{DARPA-TC} and the related work \cite{milajerdi2019poirot}, and the specific extraction results are shown in the open-source site. 
%  In order to evaluate the effectiveness of the ASG extracted by this module, we first performed Ground Truth manual extraction of attack graphs for 10 reports, i.e., we drew attack graphs that can represent their attack flow and conform to the origin graph specification based on their content descriptions. In the drawing process, we both consider the descriptions in the report \cite{DARPA-TC} and the related work \cite{milajerdi2019poirot}, and the specific extraction results are shown in the open-source site. 
We evaluated and compared the performance of ASG extraction with the latest open-source studies, AttackG and EXTRACTOR\footnote{https://github.com/ksatvat/EXTRACTOR} \cite{satvat2021extractor} from three aspects: entities, dependencies, and alignment scores. The evaluation results are shown in Table~\ref{table:asg_NodeEdgeAlignscore_eval}. The left part (columns 5 to 10) reflects the performance of the module on extraction of entity and dependency, and the right part (columns 11 to 14) integrates the performance of the module on extraction of the overall graph structure. From the left part, we find that EFI has an average improvement rate of more than \textbf{30\%} in the three comprehensive evaluation metrics (precision, recall, F-1 Score) compared with the existing work (especially, the improvement rate of recall exceeds \textbf{60\%}), which shows that our EFI is fully ahead of existing work on entity and dependency extraction. In the right part, EFI has an advantage over \textbf{200\%} in the alignment score, which indicates that the ASG extracted by EFI is closer to the ground truth in terms of the overall graph structure. In addition, although the average alignment score of EFI is only 0.593, as shown in Section~\ref{sec:6_3}, the \textit{graph alignment plus algorithm} is very sensitive to node-level modification, e.g., the alignment score drops to below 0.6 after deleting 2 nodes. And the average number of nodes in our final ASGs is 23 (\textgreater\textgreater 2), so \textbf{we believe that such deviations do not affect the graph forecast model to learn the information of graph distribution}. Finally, we chose to plot the comparison of three systems and ground truth on \textbf{TC\_A3}, as shown in Figure~\ref{fig:asg_diffsys_compare}. One may argue that the accuracy of extracted ASGs may further affect the prediction model (i.e., potential error propagation will accumulate quickly over the NLP pipeline). It is important to clarify that the role of the ASG extraction module in EFI is to enhance attack samples, with a greater focus on the integrity and rationality of these samples. To address this concern, we have employed the three-stage model from CONAN \cite{xiong2020conan} to validate all ASGs, ensuring their correct attack semantics.

\begin{figure*}[ht]
% \caption{EFI, AttackG and EXTRACTOR are compared with the extracted ASG using sample TC\_Information gather and exfiltration as input, where sub-graph A is our manually extracted Ground Truth.}
%\label{fig:asg_diffsys_compare}
\centering
\includegraphics[width=\linewidth]{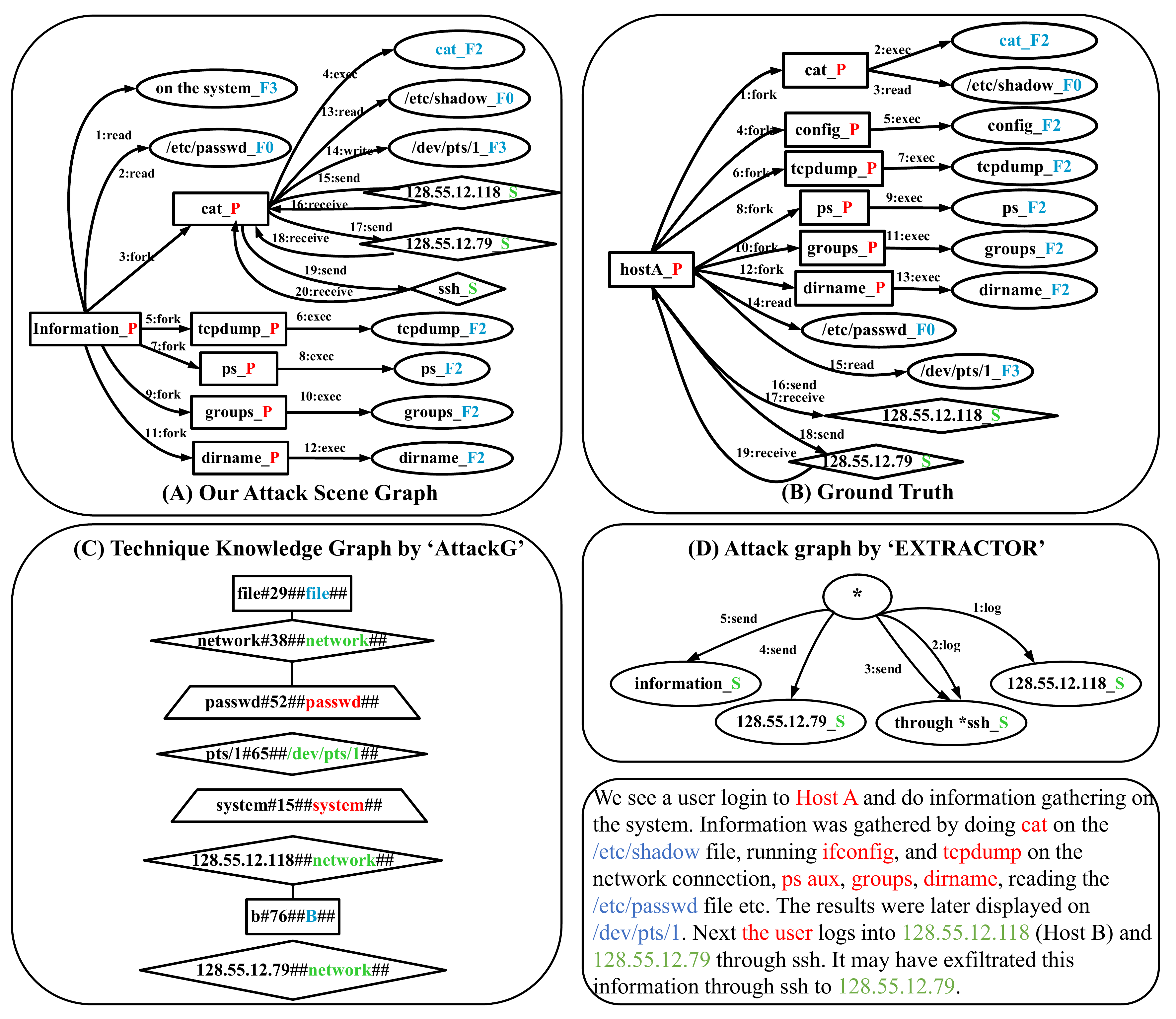}
\setlength{\abovecaptionskip}{-0.3cm}
\caption{EFI, AttackG and EXTRACTOR are compared with the extracted ASG using sample TC\_Information gather and exfiltration as input, where sub-graph A is our manually extracted Ground Truth. In the figure, red P indicates process, blue F indicates file, and green S indicates socket.}
\label{fig:asg_diffsys_compare}
\vspace{-0.2cm}
\end{figure*}

\begin{table*}[ht]
\centering
% \caption{The performance of ASG extracted by different systems in terms of entities, dependencies and alignment score. In column 1, rows 8 to 11, TC\_A1-A4 represent hc attack, ccleaner, Information gather and exfiltarion and In-memory attack with firefox in TC report, respectively. In the left half (columns 2 to 7), TP/FP/FN represent true positive/false positive/false negative, and rows 12-15 are the Precision, Recall and F-1 Score calculated by TP/FP/FN. The right half (columns 8 to 11) is the alignment score calculated by the graph alignment plus algorithm, and row 12 represents the average value of alignment scores for all graphs.
% }
\caption{The performance of technique and ASG extracted by different systems in terms of entities, dependencies and alignment score. TC\_A1-A4 represent hc attack, ccleaner attack, Information gather and exfiltarion, and In-memory attack with firefox in TC report, respectively. The right part (columns 11 to 14) is the alignment score calculated by the \textit{graph alignment plus algorithm}. }

\label{table:asg_NodeEdgeAlignscore_eval}
\scalebox{0.83}{
\begin{tabular}{|c|ccc|ccc|ccc|cccc|}
\hline
\multirow{2}{*}{\textbf{CTI Reports}} & \multicolumn{3}{c|}{\textbf{\begin{tabular}[c]{@{}c@{}}Technique (TP/FP/FN)\end{tabular}}} & \multicolumn{3}{c|}{\textbf{\begin{tabular}[c]{@{}c@{}}Entities (TP/FP/FN)\end{tabular}}} & \multicolumn{3}{c|}{\textbf{\begin{tabular}[c]{@{}c@{}}Dependencies (TP/FP/FN)\end{tabular}}} & \multicolumn{4}{c|}{\textbf{\begin{tabular}[c]{@{}c@{}}Alignment Score\end{tabular}}}                                                                                                                                                  \\ \cline{2-14} 
                                      & \multicolumn{1}{c|}{AttackG}        & \multicolumn{1}{c|}{TTPDrill}        & EFI         &\multicolumn{1}{c|}{AttackG}        & \multicolumn{1}{c|}{Extractor}        & EFI         & \multicolumn{1}{c|}{AttackG}          & \multicolumn{1}{c|}{Extractor}         & EFI            & \multicolumn{1}{c|}{\multirow{11}{*}{\textbf{}}}                                                                    & \multicolumn{1}{c|}{AttackG}                & \multicolumn{1}{c|}{Extractor}              & EFI                   \\ \cline{1-10} \cline{12-14} 
Carbanak                              & \multicolumn{1}{c|}{2/10/4}         & \multicolumn{1}{c|}{2/24/4}         & \multicolumn{1}{c|}{3/3/3}         & \multicolumn{1}{c|}{8/13/4}         & \multicolumn{1}{c|}{8/14/4}           & 10/4/2        & \multicolumn{1}{c|}{6/12/6}           & \multicolumn{1}{c|}{6/20/6}            & 11/12/2         & \multicolumn{1}{c|}{}                                                                                               & \multicolumn{1}{c|}{0.248}                  & \multicolumn{1}{c|}{0.171}                  & 0.601                  \\
DeputyDog                             & \multicolumn{1}{c|}{2/6/3}         & \multicolumn{1}{c|}{1/3/4}         & \multicolumn{1}{c|}{2/6/3}         & \multicolumn{1}{c|}{4/4/2}          & \multicolumn{1}{c|}{4/3/2}            & 5/4/1         & \multicolumn{1}{c|}{3/6/2}            & \multicolumn{1}{c|}{3/3/2}             & 4/6/1           & \multicolumn{1}{c|}{}                                                                                               & \multicolumn{1}{c|}{0.145}                  & \multicolumn{1}{c|}{0.166}                  & 0.687                  \\
DustySky                              & \multicolumn{1}{c|}{4/5/2}         & \multicolumn{1}{c|}{2/17/5}         & \multicolumn{1}{c|}{4/5/2}         & \multicolumn{1}{c|}{6/7/4}          & \multicolumn{1}{c|}{6/4/4}            & 7/6/3         & \multicolumn{1}{c|}{6/8/4}            & \multicolumn{1}{c|}{5/3/11}            & 9/10/2          & \multicolumn{1}{c|}{}                                                                                               & \multicolumn{1}{c|}{0.118}                  & \multicolumn{1}{c|}{0.019}                  & 0.422                  \\
njRAT                                 & \multicolumn{1}{c|}{4/4/3}         & \multicolumn{1}{c|}{1/22/6}         & \multicolumn{1}{c|}{4/4/3}         & \multicolumn{1}{c|}{8/10/6}         & \multicolumn{1}{c|}{6/9/8}            & 10/4/4        & \multicolumn{1}{c|}{7/23/8}           & \multicolumn{1}{c|}{5/14/10}           & 12/14/4         & \multicolumn{1}{c|}{}                                                                                               & \multicolumn{1}{c|}{0.326}                  & \multicolumn{1}{c|}{0.241}                  & 0.701                  \\
OceanLotus                            & \multicolumn{1}{c|}{3/2/1}         & \multicolumn{1}{c|}{3/44/1}         & \multicolumn{1}{c|}{3/6/1}         & \multicolumn{1}{c|}{6/11/10}        & \multicolumn{1}{c|}{5/6/11}           & 11/6/5        & \multicolumn{1}{c|}{6/11/15}          & \multicolumn{1}{c|}{4/11/17}           & 12/9/9          & \multicolumn{1}{c|}{}                                                                                               & \multicolumn{1}{c|}{0.361}                  & \multicolumn{1}{c|}{0.158}                  & 0.504                  \\
Uroburos                              & \multicolumn{1}{c|}{2/22/4}         & \multicolumn{1}{c|}{1/12/5}         & \multicolumn{1}{c|}{2/1/4}         & \multicolumn{1}{c|}{5/6/7}          & \multicolumn{1}{c|}{5/5/10}           & 8/5/4         & \multicolumn{1}{c|}{4/7/11}           & \multicolumn{1}{c|}{4/3/11}            & 7/8/8           & \multicolumn{1}{c|}{}                                                                                               & \multicolumn{1}{c|}{0.137}                  & \multicolumn{1}{c|}{0.073}                  & 0.447                  \\
TC\_A1                                 & \multicolumn{1}{c|}{3/0/3}         & \multicolumn{1}{c|}{1/10/8}         & \multicolumn{1}{c|}{6/4/3}         & \multicolumn{1}{c|}{5/3/8}          & \multicolumn{1}{c|}{8/11/8}           & 11/3/2        & \multicolumn{1}{c|}{4/4/12}           & \multicolumn{1}{c|}{6/7/13}            & 11/14/5         & \multicolumn{1}{c|}{}                                                                                               & \multicolumn{1}{c|}{0.212}                  & \multicolumn{1}{c|}{0.167}                  & 0.514                  \\
TC\_A2                                 & \multicolumn{1}{c|}{5/11/2}         & \multicolumn{1}{c|}{3/7/3}         & \multicolumn{1}{c|}{6/3/1}         & \multicolumn{1}{c|}{4/1/8}          & \multicolumn{1}{c|}{7/10/5}           & 10/7/2        & \multicolumn{1}{c|}{3/3/10}           & \multicolumn{1}{c|}{6/6/7}             & 11/8/2          & \multicolumn{1}{c|}{}                                                                                               & \multicolumn{1}{c|}{0.085}                  & \multicolumn{1}{c|}{0.025}                  & 0.672                  \\
TC\_A3                                 & \multicolumn{1}{c|}{1/0/7}         & \multicolumn{1}{c|}{0/6/8}         & \multicolumn{1}{c|}{4/6/4}         & \multicolumn{1}{c|}{3/5/15}         & \multicolumn{1}{c|}{4/5/14}           & 16/2/2        & \multicolumn{1}{c|}{2/3/17}           & \multicolumn{1}{c|}{3/4/16}            & 17/6/2          & \multicolumn{1}{c|}{}                                                                                               & \multicolumn{1}{c|}{0.167}                  & \multicolumn{1}{c|}{0.194}                  & 0.833                  \\
TC\_A4                                 & \multicolumn{1}{c|}{2/10/4}         & \multicolumn{1}{c|}{1/16/5}         & \multicolumn{1}{c|}{3/2/1}         & \multicolumn{1}{c|}{5/4/4}          & \multicolumn{1}{c|}{5/10/4}           & 7/4/2         & \multicolumn{1}{c|}{3/5/8}            & \multicolumn{1}{c|}{4/8/7}             & 10/9/1          & \multicolumn{1}{c|}{}                                                                                               & \multicolumn{1}{c|}{0.249}                  & \multicolumn{1}{c|}{0.183}                  & 0.545                  \\ \hline
\textbf{Overall Precision}            & \multicolumn{1}{c|}{0.46}          & \multicolumn{1}{c|}{0.11}          & \multicolumn{1}{c|}{0.51}          & \multicolumn{1}{c|}{0.495}          & \multicolumn{1}{c|}{0.450}            & \textbf{0.668}         & \multicolumn{1}{c|}{0.382}            & \multicolumn{1}{c|}{0.418}             & \textbf{0.513}           & \multicolumn{1}{c|}{\multirow{3}{*}{\textbf{\begin{tabular}[c]{@{}c@{}}Overall\\ Alignment \\ Score\end{tabular}}}} & \multicolumn{1}{c|}{\multirow{3}{*}{0.205}} & \multicolumn{1}{c|}{\multirow{3}{*}{0.140}} & \multirow{3}{*}{\textbf{0.593}} \\
\textbf{Overall Recall}               & \multicolumn{1}{c|}{0.42}          & \multicolumn{1}{c|}{0.23}          & \multicolumn{1}{c|}{0.59}          & \multicolumn{1}{c|}{0.474}          & \multicolumn{1}{c|}{0.478}            & \textbf{0.778}         & \multicolumn{1}{c|}{0.358}            & \multicolumn{1}{c|}{0.350}             & \textbf{0.759}           & \multicolumn{1}{c|}{}                                                                                               & \multicolumn{1}{c|}{}                       & \multicolumn{1}{c|}{}                       &                        \\
\textbf{Overall F-1 Score}            & \multicolumn{1}{c|}{0.43}          & \multicolumn{1}{c|}{0.15}          & \multicolumn{1}{c|}{0.53}          & \multicolumn{1}{c|}{0.461}          & \multicolumn{1}{c|}{0.452}            & \textbf{0.715}         & \multicolumn{1}{c|}{0.339}            & \multicolumn{1}{c|}{0.359}             & \textbf{0.605}           & \multicolumn{1}{c|}{}                                                                                               & \multicolumn{1}{c|}{}                       & \multicolumn{1}{c|}{}                       &                        \\ \hline
\end{tabular}
}
\vspace{-0.2cm}
\end{table*}

\begin{figure*}[ht]
% \caption{The performance of forecast model on the validation set with the epoch number of training, i.e., the loss value and TPR of Node and Edge predictions on the validation set, where the TPR of edges is calculated considering non-zero edges only.}
%\label{fig:genmodel_loss_TPR}
\centering
\subfigure[M=5]{\includegraphics[width=8cm]{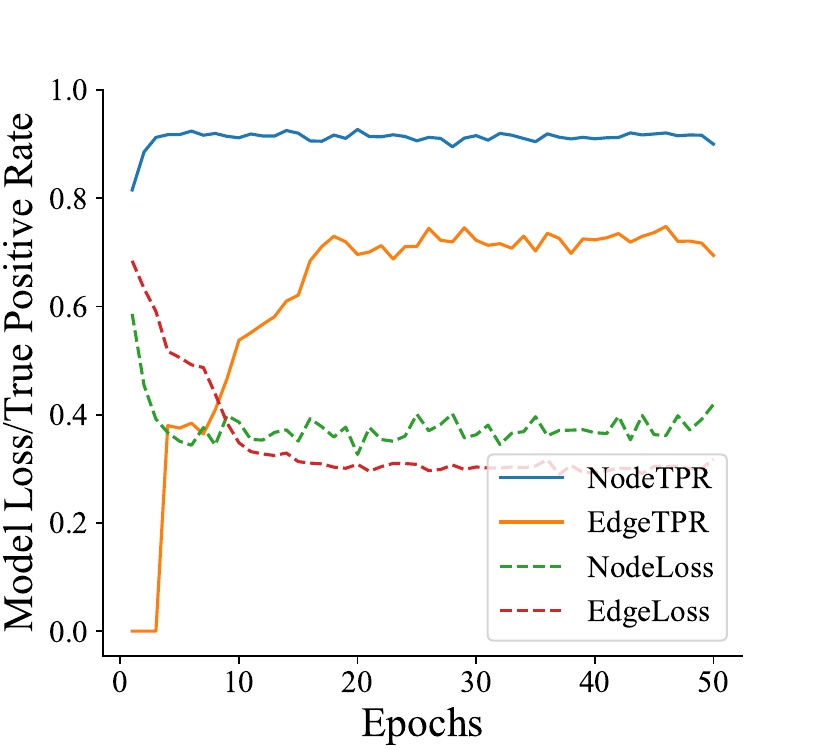}\label{fig:genmodel_loss_TPR_left}}
\subfigure[M=28]{\includegraphics[width=8cm]{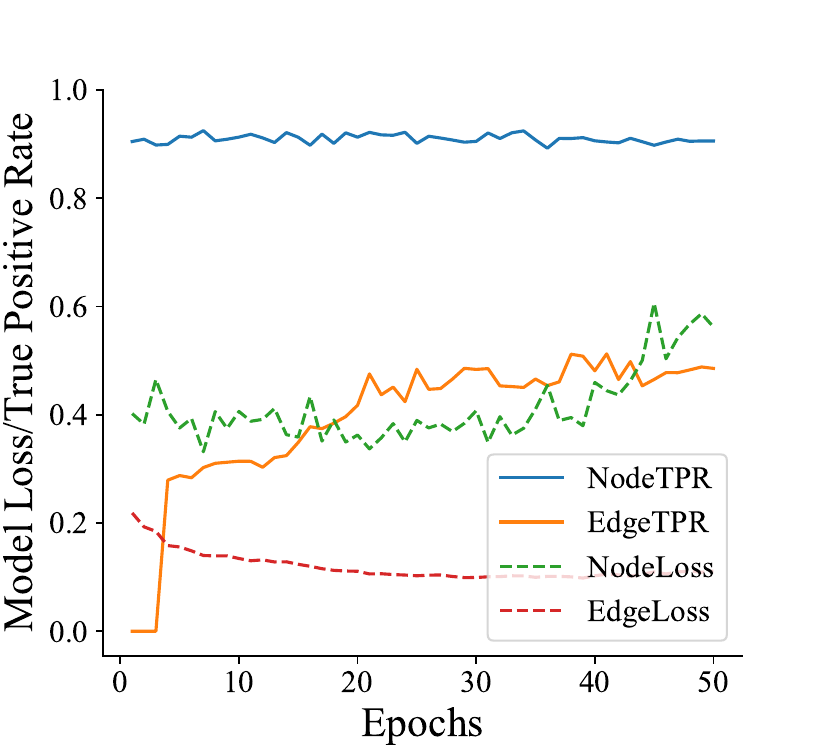}\label{fig:genmodel_loss_TPR_right}}
% \setlength{\abovecaptionskip}{-0.2cm}
% \caption{The performance of forecast model on the validation set with the epoch number of training, i.e., the loss value and TPR of Node and Edge predictions on the validation set, where the TPR of edges is calculated considering non-zero edges only.}
\caption{The performance of forecast model on the test set. The EdgeTPR is calculated by non-zero edges.}
\label{fig:genmodel_loss_TPR}
\vspace{-0.5cm}
\end{figure*}

\subsection{RQ4: How to prove the effectiveness of AFG generation?}
\label{sec:6_5}
\par
We need to state that, as shown in Section~\ref{sec:4_2}, the training set of graph forecast model is ASGs (1,429 in total) extracted automatically from CTI reports, and both the input and output of the model are abstract heterogeneous graphs without node names. In order to answer R4, we evaluate the AFG generation of EFI from three aspects: training performance, ASG reconstruction, forecast and interpretation performance and APG reconstruction. 
% In the process of applying graph forecast model, we would input a graph with N nodes, the model predicts the attribute of next node(N+1) and the dependency between this node and the previous N nodes as the output. Then we would merge output and input in this iteration as the input with N+1 nodes in next iteration.
% In order to answer R4, we first need to state that, as shown in the previous Section~\ref{sec:4.2}, the training set of the graph forecast model is ASGs (1911 in total) extracted automatically from a large number of open-source CTI reports, and both the input and output of the model are abstracted heterogeneous graphs, i.e., the heterogeneous graphs all have only node attributes and edge attributes without node naming. And in the process of graph forecast model application, we would input the heterogeneous graph (total N nodes), the model predicts the node properties of the next node N+1 of this graph and the edge relationship between this node and the previous N nodes as the output, and then we would merge this output and input as the next input (total N+1 nodes), traversing until the predicted node is empty or the set prediction length.
\subsubsection{Training Performance}
\label{sec:6_5_1}
We divided the training set and test set according to the ratio of 8:2 (each ASG is in article level), and recorded metric after each epoch, as shown in Figure~\ref{fig:genmodel_loss_TPR_left}. This sub-graph represents the evaluation performance of the model on the test set. For the specific evaluation metrics we chose the node-level and edge-level loss, as well as the true positive rate (TPR). After more than 40 iterations, \textbf{the TPRs of node attributes and edge attributes reach 92\% and 73\%, respectively}. The main reason why the TPR of edge attributes is lower than that of node is that the adjacency matrix of the graph is sparse (less than 100 non-zero values exist in a 40x40 adjacency matrix), which leads to a sample imbalance problem (the edges with label 0 are much larger than the other edge types, i.e., no such edge exists). We reduced the impact of this problem by limiting the number $M$ of forward prediction, and we set $M$ to 5 and 28 based on the mean and maximum values of the temporal differences between subjects and objects present in all edges within our dataset. We designed control experiments under these two $M$, as shown in Figure~\ref{fig:genmodel_loss_TPR}.  Finally, it can be observed in Figure~\ref{fig:genmodel_loss_TPR_right} that the node-level loss cannot converge properly, and the edge-level prediction accuracy ends up at only 48\%. We believe that the sparsity in the adjacency matrix brings a large impact on the forecast accuracy of edges, and our countermeasure effectively mitigates this problem.
\subsubsection{ASG Reconstruction}
\label{sec:6_5_2}
We apply the graph forecast model to the ASG reconstruction to illustrate the effectiveness of AFG generation (i.e., \textbf{simulating the forecast with ASG reconstruction}). The schematic diagram of the experimental procedure is shown in Appendix~\ref{sec:appendix_approach_evalGFM}. First, we traversed and interpreted all the ASGs (also called original ASG) of test set in technique-level (Figure~\ref{fig:Generation_Model_Evaluation_Method_Design_a}, T1566.001, T1204.002, T1574). Second, we deleted nodes (the maximum number of deletion is 5) and corresponding edges from the original ASG based on their temporal relevance in a post-order traversal until the deleted subgraph satisfied an ATG, resulting in a broken ASG. We calculated the alignment score between the broken ASG and the original ASG, and recorded the corresponding list of ATGs associated with the broken ASG (Figure~\ref{fig:Generation_Model_Evaluation_Method_Design_b}, T1566.001, T1204.002). Then, we input the broken ASG into the trained graph forecast model and generated nodes and edges iteratively until the number of ATGs corresponding to the generated AFG exceeded that of the broken ASG. \textbf{We refer to this process as ASG reconstruction}. Finally, we recorded the number of generated nodes and edges, the alignment score of the generated AFG (reconstructed ASG) and the original ASG, and the list of ATGs corresponding to the generated AFG (Figure~\ref{fig:Generation_Model_Evaluation_Method_Design_c}, T1566.001,T1204.002,T1574). 
\par
After traversing all ASGs of test set, we collected alignment scores under different scenarios as shown in Figure~\ref{fig:alignsocre_afg_brokenasg_left}. We can see that the alignment score between the broken ASG and the original ASG (blue line in Figure~\ref{fig:alignsocre_afg_brokenasg_left}) decreases very fast (only 0.067 when N=5), which is due to the sensitivity of the \textit{graph alignment plus algorithm} to the node, as shown in Section~\ref{sec:6_3}. 
%  Besides, It can be seen that the overall alignment score gradually decreases as the number of deleted nodes increases, similar to the performance in Figure~\ref{fig:graphalign+}. 
Besides, the average alignment scores of AFG under each number of nodes are much higher than those of broken ASGs (\textbf{exceeding 0.27 on average and reaching up to 0.32 when N=5}), indicating that graph forecast model can effectively help the broken ASG recover to the original ASG.

\subsubsection{APG Reconstruction}
\label{sec:6_5_3}
In the real scenario, EFI requires APGs provided by EDR tools as input to make forecasts in syslog-level and perform strategy assignment. Therefore, \textbf{we also tested the effectiveness of EFI for APG reconstruction as a way to verify the effectiveness of EFI for AFG generation in real scenarios.} We reproduced DEPIMPACT~\cite{fang2022back} for corelated sub-graph investigation of alert point, and reproduced DEPCOMM~\cite{xu2022depcomm} and CPR~\cite{xu2016cpr} for compressing redundant edges and nodes in the sub-graph, to obtain APGs constructed by EDR tools in real scenarios. We chose four attacks (TC\_A1-A4) from DARPA Engagement and finally obtained four APGs.
\par
Similar to ASG reconstruction, we started by constructing the broken APG and then proceeded to reconstruct a new AFG. We recorded alignment scores of the broken APG and the generated AFG from the broken APG under different numbers of nodes generation/deletion, as shown in Figure~\ref{fig:alignsocre_afg_brokenasg_right}. First, we can find that the effect of node deletion on APGs is smaller than that on ASGs (shown by the blue line in Figure~\ref{fig:alignsocre_afg_brokenasg_left} and Figure~\ref{fig:alignsocre_afg_brokenasg_right}), the reason is that the average number of nodes (18) and edges (21) in the APG is greater than the average number of nodes (11.3) and edges (10.8) in the ASG. In addition, the average alignment score of AFGs are higher than that of the broken APG, exceeding 0.24 on average and 0.27 at maximum. It also indicates that the graph prediction model can effectively assist in recovering the broken APG back to its original state. Due to the intricate structure of the APG, the average increase in alignment score of the reconstructed APG (0.23) is smaller than that of the reconstructed ASG (0.27). However, the goal of EFI is to predict the next move in syslog-level and the alignment score of the reconstructed APG reaches 0.8 when N=1, which indicates that the graph forecast component of EFI will perform well in realistic scenarios in cooperation with EDR tools.

\subsubsection{Forecast and Interpretation Performance}
\label{sec:6_5_4}
In addition to focusing on the evolution of alignment scores, we also make Table~\ref{table:afg_NodeEdgeTech_eval} to illustrate the forecast performance of EFI in technique-level. From Table~\ref{table:afg_NodeEdgeTech_eval}, we can see that the average deletion of 2.4 nodes decreases the number of ATGs by 4, and the average generation of 1.7 nodes increases the number of ATGs by 3.5. The reason is that it is easy to match multiple ATGs with similar structures after the graph structure is increased.

The metrics in the fifth to seventh rows are calculated using the list of ATGs from original ASG as the ground truth. The broken ASG is obtained by deleting nodes in a post-order traversal of the original ASG, so the list of ATGs from broken ASG must be part of that from original ASG (i.e., precision=1). As we can see, the list of ATGs from AFG performed extremely well on all three metrics, exceeding 90\%.

% However, the ultimate purpose of EFI is to perform strategy dispatch for advance reinforcement through predictive techniques, so the variation in alignment scores alone is not sufficient to fully represent the value of the system for use in real scenarios. Therefore, we plotted the Table~\ref{table:afg_NodeEdgeTech_eval} on the forecast performance of EFI in technique-level. From the Table~\ref{table:afg_NodeEdgeTech_eval}, we can see that the average deletion of 2.4 nodes decreases the technique templates by 4, and the average generation of 1.7 nodes increases the technique templates by 3.5. This is because it is easy to match multiple technique templates with similar structures after the graph structure is increased. But since EFI attempts dispatch strategies to  EDR for reinforcement, we believe that reducing false negatives is the key. In addition, the metrics in the fifth to seventh rows are calculated using the technique template list of original ASG  as the ground truth, and the broken ASG is obtained by deleting nodes in a post-order traversal of the original ASG, so its technique template list must be part of the original ASG's , i.e., Precision=1. As we can see, the technique template list of AFG  performed extremely well in all three metrics, exceeding 90\%, so we believe that EFI is extremely effective and usable in the real scenario.
\par
\textbf{Overhead:} Since attackers tend to prioritize lateral movement to the target host or hide themselves after invading the environment. We evaluate the time, CPU and memory overhead, as well as the real-time requirement for each component of EFI, as shown in Table~\ref{table:overhead}. It can be observed that the average time to extract the ASG is 72.7s and the time to process all CTI reports (3,484) is over 253,000 seconds, but we have no real-time requirement for this component. The average time to generate AFG is less than 4s and the average time to match individual ATG is less than 2s. Even for interpreting a whole AFG, the total time is only 5 mins, which is much less than the average time for lateral movement(1h 58mins) \cite{lateral-movement}. EFI also has a low overhead in memory (less than 400MB) and CPU (around 15\%), and it is entirely possible to reduce the time to process multiple alerts issued by the EDR and explain the whole AFG by multi-processes.
% At the same time, EFI will automatically dispatch strategies to EDR for advance reinforcement based on the interpretation results, thus preventing post-exploitation attacks before they occur and greatly reducing MTTR.
% Finally, the use of EFI in real scenarios is to automatically generate and interpret AFGs based on the APGs provided by EDR, and then to dispatch strategies to EDR for advance reinforcement. This process has a high requirement for real-time, because the attacker tends to prioritize lateral movement to the target host or hide itself after invading the environment, and the average time for lateral movement is 1 h 58 min \cite{lateral-movement}.Therefore, we evaluate the time overhead of EFI in AFG generation and interpretation, as shown in Table \ref{table:overhead}.  We can see that the average time for generating AFG is less than 4s and matching individual ATG is less than 2s, and even for interpreting a whole AFG, the total time is only 5 minutes, which is much less than the average time for lateral movement. At the same time, EFI will automatically dispatch strategies to EDR for advance reinforcement based on the interpretation results, thus preventing problems before they occur and greatly reducing MTTR (Mean-time-to-response).

\begin{table}[ht]
% \caption{Experimental results of EFI evaluation on graph prediction and interpretation. Rows 2-4 are the average number of nodes, edges and technique templates, and rows 5-7 are the Precision/Recall/F-1 score obtained by calculating the TP/FP/FN of the technique templates using the original ASG technique template as the ground truth.}
\caption{Results of graph forecast and interpretation. Rows 2-4 are the average number of nodes, edges and technique templates, and rows 5-7 are the precision/recall/F-1 score obtained by calculating the TP/FP/FN using the ATGs corresponding to original ASG as the ground truth.}
\label{table:afg_NodeEdgeTech_eval}
\centering
\scalebox{1}{
\begin{tabular}{|c|c|c|c|}
\hline
                           & \textbf{\begin{tabular}[c]{@{}c@{}}Original \\ ASG\end{tabular}} & \textbf{\begin{tabular}[c]{@{}c@{}}Broken\\ ASG\end{tabular}} & \textbf{AFG}   \\ \hline
Node                       & 11.3                                                             & 8.9                                                           & 10.6           \\ \hline
Edge                       & 10.8                                                             & 8.1                                                           & 9.9            \\ \hline
The number of ATGs         & 13.5                                                             & 9.5                                                           & 13             \\ \hline
\textbf{Overall Precision} & 1                                                                & 1                                                             & \textbf{0.918} \\ \hline
\textbf{Overall Recall}    & 1                                                                & 0.641                                                         & \textbf{0.903} \\ \hline
\textbf{Overall F1-socre}  & 1                                                                & 0.717                                                         & \textbf{0.906} \\ \hline
\end{tabular}
}
% \vspace{-0.3cm}
\end{table}

\begin{figure*}[ht]
% \caption{Change in graph alignment score with graph structure modification.A negative number in the horizontal coordinate indicates deletion and a positive number indicates an increase.}
%\label{fig:alignsocre_afg_brokenasg}
\centering
\setlength{\abovecaptionskip}{-0.15cm}
\subfigure{\includegraphics[width=8cm]{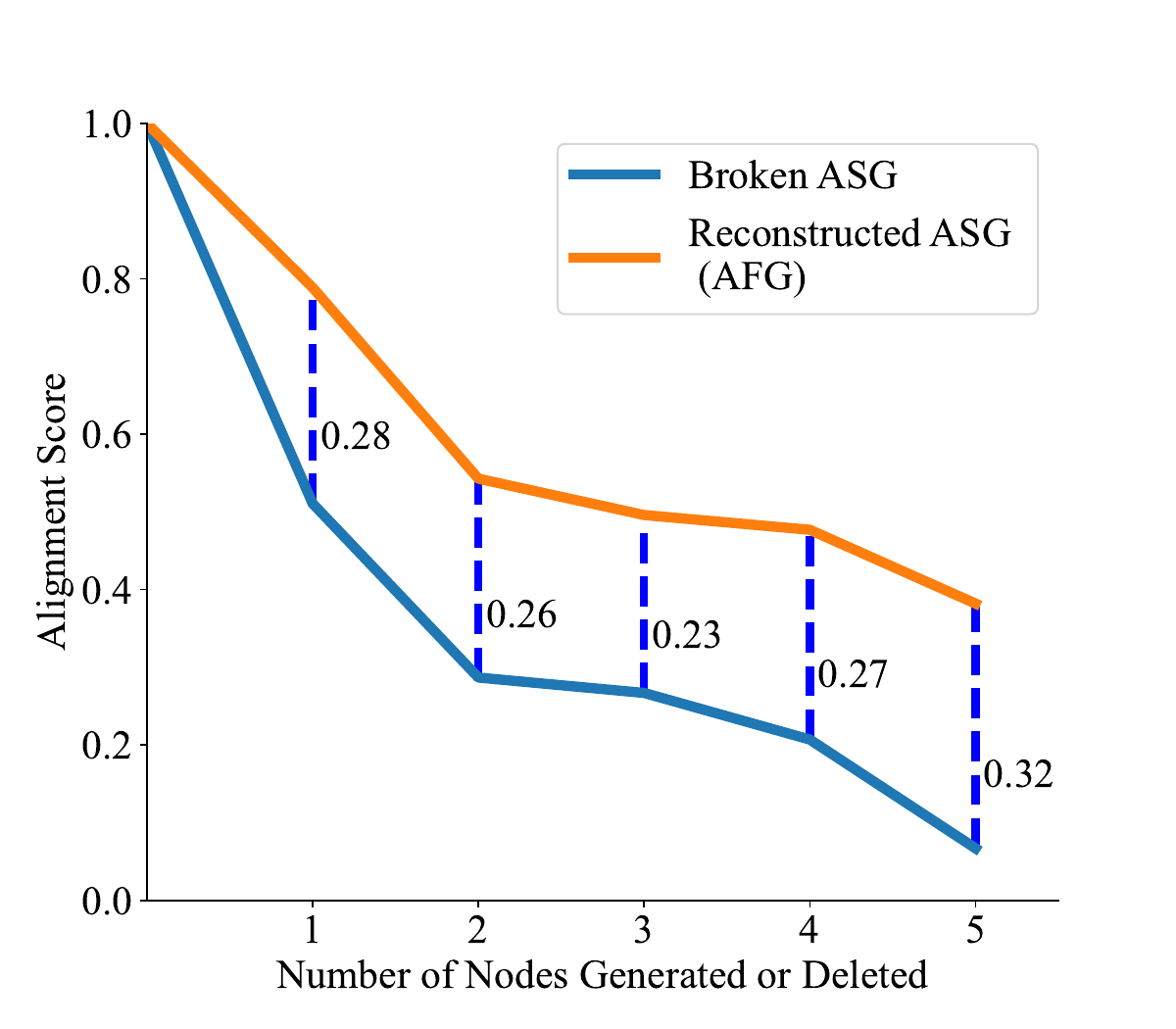}\label{fig:alignsocre_afg_brokenasg_left}}
\subfigure{\includegraphics[width=8cm]{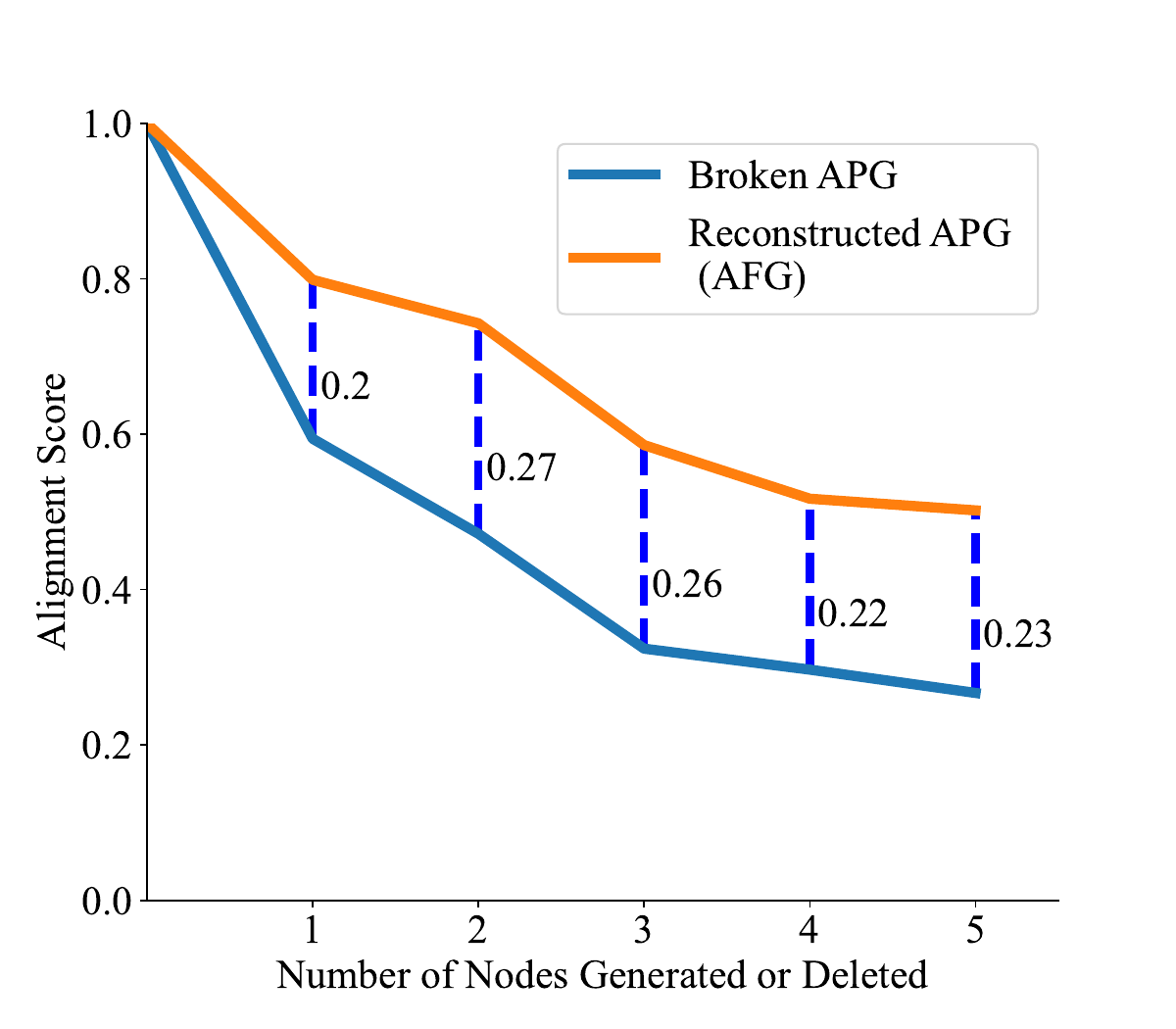}\label{fig:alignsocre_afg_brokenasg_right}}
\caption{Change in graph alignment score with graph structure modification. The horizontal coordinate value is the number of node deletions for broken ASG/APG and the number of node generation for AFG. The vertical dashed line indicates the difference in graph alignment score between the broken graph and the AFG.}
\label{fig:alignsocre_afg_brokenasg}
\vspace{-0.2cm}
\end{figure*}

\begin{table*}[ht]
\caption{The overhead and real-time requirements of the EFI component. The second column is the overhead of extracting ASG, the third column is the overhead of generating AFG, the fourth column is the overhead of investigating APG (from \cite{fang2022back,xu2022depcomm}), the fifth column is the overhead of computing individual ATG alignment scores, and the sixth column is the overhead of interpreting a whole AFG.}
\label{table:overhead}
\centering
\begin{tabular}{|c|c|c|c|c|c|}
\hline
\textbf{Overhead}                                                         & \textbf{\begin{tabular}[c]{@{}c@{}}ASG \\ Extraction\end{tabular}} & \textbf{\begin{tabular}[c]{@{}c@{}}AFG\\ Generation\end{tabular}} & \textbf{\begin{tabular}[c]{@{}c@{}}APG\\ Investigation\end{tabular}} & \textbf{\begin{tabular}[c]{@{}c@{}}One ATG\\ Match\end{tabular}} & \textbf{\begin{tabular}[c]{@{}c@{}}Whole AFG\\ Interpretation\end{tabular}} \\ \hline
\textbf{Time}                                                             & 72.7s                                                              & 3.6s                                                              & 314.8s                                                               & 1.9s                                                              & 307.9s                                                              \\ \hline
\textbf{CPU}                                                              & 11.2\%                                                              & 7.1\%                                                              & 14.9\%                                                              & 6.5\%                                                              & 6.7\%                                                                \\ \hline
\textbf{\begin{tabular}[c]{@{}c@{}}Real-time \\ Requirement\end{tabular}} & No                                                                 & Yes                                                               & Yes                                                               & Yes                                                               & Yes                                                                  \\ \hline
\textbf{Memories}                                                         & 337.4MB                                                            & 281.1MB                                                           & 376.0MB                                                           & 169.7MB                                                           & 170.1MB                                                              \\ \hline
\end{tabular}
\vspace{-0.2cm}
\end{table*}

% \begin{figure*}[h!t]
% % \caption{The performance of forecast model on the validation set with the epoch number of training, i.e., the loss value and TPR of Node and Edge predictions on the validation set, where the TPR of edges is calculated considering non-zero edges only.}
% %\label{fig:genmodel_loss_TPR}
% \centering
% \subfigure[M=5]{\includegraphics[width=8cm]{Fig/Generation Model Performance on Val Dataset.png}\label{fig: loss_left}}
% \subfigure[M=28]{\includegraphics[width=8cm]{Fig/Generation Model Performance on Val Dataset M28.png}\label{fig: loss_right}}
% \setlength{\abovecaptionskip}{-0.1cm}
% \caption{The performance of forecast model on the validation set with the epoch number of training, i.e., the loss value and TPR of Node and Edge predictions on the validation set, where the TPR of edges is calculated considering non-zero edges only.}
% \label{fig:genmodel_loss_TPR}
% \end{figure*}
\section{Discussion and Limitation}\label{sec:discuss}
\par
% \textbf{ASG extraction:} Regarding the automatic extraction of ASGs from the CTI reports, as shown in table \ref{table:asg_NodeEdgeAlignscore_eval}, we notice that there are irrelevant edges or nodes (missing relevant edges or nodes) in ASGs, i.e., FP and FN. Because CTI combines two features: natural language complexity (large gap between different authors' styles) and prominent domain features (cyber security terminology). However, it is important to note that the purpose of EFI to extract ASGs representing attacks from a large number of open source CTI reports is to train graph forecast model, which are more biased towards learning the overall graph  distribution than the details. In addition, we abstract the nodes to further weaken the impact of less-than-perfect ASG extraction on model training. \textbf{Future Work}, we can perform text clustering on CTI reports and perform template extraction on their corresponding ASGs based on the clustering results (i.e., one type of template corresponds to one type of APT attack), and then generate more and more accurate ASGs based on the templates.
\textbf{ASG extraction:} Regarding the automatic extraction of ASGs from the CTI reports, as shown in Table~\ref{table:asg_NodeEdgeAlignscore_eval}, we notice that there are several FPs and FNs. The reason is that CTI reports combine two features: natural language complexity and prominent domain features. Fortunately, our graph forecast model is based on graph distribution rather than graph details. In addition, we abstract the nodes to further weaken the impact of less-than-perfect ASG extraction on model training. In the future, we can perform text clustering and template extraction on CTI reports to generate  more accurate ASGs.
\par
% \textbf{Graph alignment plus algorithm:} Regarding the graph alignment plus algorithm, its performance is shown in figure \ref{fig:graphalign+} and table \ref{table:asg_tech_eval}. But its time overhead is large, the time complexity can reach $O(N^4)$ for two N-node inputs, which is due to the existence of multiple layers of nesting in our designed algorithm to complete the candidate nodes and fixed nodes for the purpose. However, in the real running scenario of EFI is only used to interpreter AFG, and we found by statistics that the average number of nodes of ASG is 30.4 and the average number of nodes of ATG is 5.1. The average time to determine whether a ATG exists in a AFG is less than 2s, and the average time to traverse all ATGs is only 5 minutes, as shown in table \ref{table:time_overhead}, which is sufficient for  relative real-time to the average lateral movement time (1 h 58 min).
% \textbf{Graph alignment plus algorithm:} The time complexity of our \textit{graph alignment plus algorithm} can reach $O(N^4)$ for two N-node inputs because the existence of multiple layers of nesting. However, in the real-world usage scenario, the average number of nodes of ASG is 30.4 and the average number of nodes of ATG is 5.1, and the average time to traverse all ATGs is only 5 minutes. It is sufficient for relative real-time to the average lateral movement time (1 h 58 min).
\par
% \textbf{ATG and graph forecast model:} For the ATG we extracted manually from the atomic red team \cite{atomic-red-team}, due to the nature of the atomic technique, the number of nodes and edges per ATG is small (only 5.1 nodes and 4.7 edges), which leads to the graph alignment plus algorithm often has a high FP when using a ATG to interpret a AFG. However, we care more about low FN because the reinforcement does not affect the normal environment, but the false negatives can lead to financial losses of attackers. In addition, the atomic technique covers only 12 tactics and 123 techniques, and is missing 2 tactics and 68 techniques. \textbf{Future Work}We believe that the node and edge attributes can be subdivided again, e.g., subdividing the registry into HKC,HKLM,HKU, etc., thus alleviating the problem of too much similarity between atomic technologies. While for the graph forecast model, it can't generate a AFG with the number of matching ATGs plus one within a predetermined M steps (M=5) for some inputs. We consider that these inputs already have a complete structure on attack graph and thus cannot continue to be predicted.
\textbf{ATG and graph forecast model:} Due to the nature of the atomic technique, the average number of nodes and edges per ATG is small (only 5.1 nodes and 4.7 edges), which leads to some FPs when using a ATG to interpret a AFG. In the future, we will also try to subdivide node and edge attributes (e.g., subdividing the registry into HKC, HKLM, and HKU) to mitigate the problem. While for the graph forecast model, it can't generate a AFG with the number of matching ATGs plus one within a predetermined M steps (M=5) for some inputs. We consider that these inputs already have a complete structure on attack graph and thus cannot continue to be predicted.
\section{Related Work}\label{sec:related}
\par 
% \textbf{Information extraction from CTI:} To keep up with the rapidly evolving forms of attacks, cyber threat intelligence plays an important supporting role in security confrontations \cite{gao2021enabling, milajerdi2019holmes}. There are already many open standard formats (e.g., OpenIOC \cite{OpenIOC}, STIX \cite{STIX}) in the cyber security community to use IOC as an important tool to resist attacks. For example, OpenIOC shares MD5 value of the malware on the platform for vendors to use as a base for detection. The data shared by these open source platforms, while structured and machine-readable, is single-point independent and lacks information about interactions with other IOCs, which leads to the inability of vendors to capture complete attacks. In addition attackers can evade detection by changing IOC information frequently because its cost is very low (e.g., changing IPs, domain names, software feature values).
\textbf{Information extraction from CTI:} CTI plays an important role in security confrontations \cite{gao2021enabling, milajerdi2019holmes}. The data shared by these open source platforms, while structured and machine-readable, is single-point independent and lacks information about interactions with other IOCs, which leads to the inability of vendors to capture complete attacks.
\par
Recently researchers have attempted to automate the extraction of information from unstructured CTI reports. iACE \cite{liao2016acing} proposes a graph mining technique to collect IOCs from CTI reports, improving the recognition rate of IOCs and NERs, but still missing the classification of inter-entity dependencies. 
TTPDrill \cite{husari2017ttpdrill} uses a dependency interpreter to extract binary groups (verbs, objects) from each sentence and determine the attack technique corresponding to this sentence by calculating the similarity. However, this work does not analyze the contextual information of the sentence, while relying extremely on the benchmark dataset (called Ontology in its work). 
ChainSmith~\cite{zhu2018chainsmith} extracts IOCs using regular expressions and maps them to a certain campaign phase using a classifier. However, they ignore the relationship between IOCs and do not consider other attack-related entities (as shown in Section~\ref{sec:4_1}) that play an important role in CTI reports.
EXTRACTOR \cite{satvat2021extractor} and ThreatRaptor \cite{gao2021enabling} extract information from CTI reports through a custom NLP pipeline, but both suffer from certain problems. EXTRACTOR classifies all IOC-free entities as the same malicious entity, resulting in a star-shaped structure of the generated graph while ignoring the case where the attacker uses multiple malicious processes to reach one target. ThreatRaptor extracts relationships between IOCs to build the traceability graph, but ignores common co-reference cases in CTI reports. 
AttackG \cite{li2021attackg} tends to extract the Technique Knowledge Graph, but it lacks information extraction of specific types of dependencies between entities and does not perform text redundancy filtering, which lead to the generation of many separate sub-graphs. Unlike previous work, the ASG extracted by EFI is a holistic heterogeneous graph with node and dependency attributes that can be mapped to the low-level system logs.

\par
\textbf{Technique-level interpretation:} With the increasing usage of system logs for causal analysis and multi-stage attack detection \cite{bates2015trustworthy, king2003backtracking, pohly2012hi}, analysts have found a huge semantic gap between the low-level system logs and the high-level techniques, i.e.,  a large combination of low-level logs is required to reflect the high-level technique. Therefore, Analysts attempt to combine provenance graph with ATT\&CK to map low-level system logs to high-level attackers' technique aid in a comprehensive analysis. Both HOLMES \cite{milajerdi2019holmes} and RapSheet \cite{hassan2020tactical} design detection rules by observing how the technique is implemented and determine the technique stage corresponding to the alert stage. However, they both rely entirely on manual policies, making it difficult to identify novel attack techniques. To interpret the attack graphs extracted from CTI reports in technique-level, Poirot \cite{milajerdi2019poirot} designed a graph alignment algorithm that predetermines candidate nodes and calculates contribution scores separately to fix structure to find possible attack graphs embedded in provenance graph. However, the extraction of attack graphs is manual and does not consider the multi-hop equivalent semantic case which is common in attacks. AttackG \cite{li2021attackg} implements the automatic extraction of technique templates and attack graphs, and interprets the attack graphs at a technique-level. However, technical templates and attack graphs generated by AttackG are discrete, leading further generalized impossible. In addition, AttackG focuses too much on the names of entities and leads to a poor graph alignment score. On the contrary, we combine elaborate ATGs and \textit{graph alignment plus algorithm}, consider abstract nodes and multi-hop semantic equivalence cases, to interpret AFG and minimize false negatives.
\par
\textbf{Generation algorithms and models:} There have been a number of studies on deep learning generative models, but all of them have certain problems. For example, VGAE \cite{kipf2016vgae} and Graphite \cite{grover2019graphite} propose graph generation models based on variational autocoders, but they are limited to learning from a single graph. GraphVAE \cite{simonovsky2018graphvae} propose a VAE-based approach to generate graphs, while it only can generate small graphs with less than 40 nodes. In addition, GraphRNN \cite{you2018graphrnn} proposes a neural network for sequentially generating graph to satisfy the requirement of large graphs, but it only considered isomorphic graphs. Both Tiresias~\cite{shen2018tiresias} and DEEPCASE \cite{van2022deepcase} utilize the alert sequences of EDR and combine them with a neural network to predict the next alert. However, their predictive capabilities are limited to EDR (Symantec) alert levels for non-graphical data structures, thus lacking universality. Unlike previous work, EFI's graph forecast model can automatically learn graph distributions from massive sample graphs and generate heterogeneous graphs with node attributes and dependency attributes of arbitrary size.
\section{Conclusion}\label{sec:conclusion}
% In this paper, we propose a real-time post-exploitation attack forecasting and interpreting system, EFI. We first automate the extraction of ASG, which can be mapped to low-level system logs, from a large number of open-source CTI reports to train our graph forecast model. Then we use APG provided by EDR as the input to the model, and make the output as our AFG. Finally we use mannully constructed ATG and \textit{graph alignment plus algorithm} to interpret the AFG on technique-level, so as to predict the attacker's next possible technique. EFI then automatically dispatches strategies to EDR based on interpretation results and reinforces in advance, while avoids the impact of EDR's false positive. Because even if it does occur, the reinforcement does not affect the environment of enterprise. 

In this paper, we propose a real-time post-exploitation attack forecasting and interpreting system, EFI. It can get one step ahead of the attacker, automatically predict next move during post-exploitation, explain attacks in technique-level and dispatch strategies to EDR for advance reinforcement. It also can avoid the impact of existing EDR false positives, and reduce the attack surface of system without affecting the normal operations. Experiments show that the forecast and interpretation precision of EFI can reach 91.8\%. To the best of our knowledge, we are the first to work on automated forecasting and interpreting post-exploitation attacks in syslog-level.

\bibliographystyle{IEEEtran}
\bibliography{main}

\clearpage
\appendices
\setcounter{equation}{0}
\pagestyle{empty}

% \appendices
% \section{Appendix}
\label{appendix}

\begin{figure*}[bp]
% \caption{The examples of manually drawn attack template graph(ATG).}
% \label{fig:technique template schematic}
\centering
\subfigure[T1119]
{
    \begin{minipage}[b]{.45\linewidth}
        \includegraphics[scale=0.33]{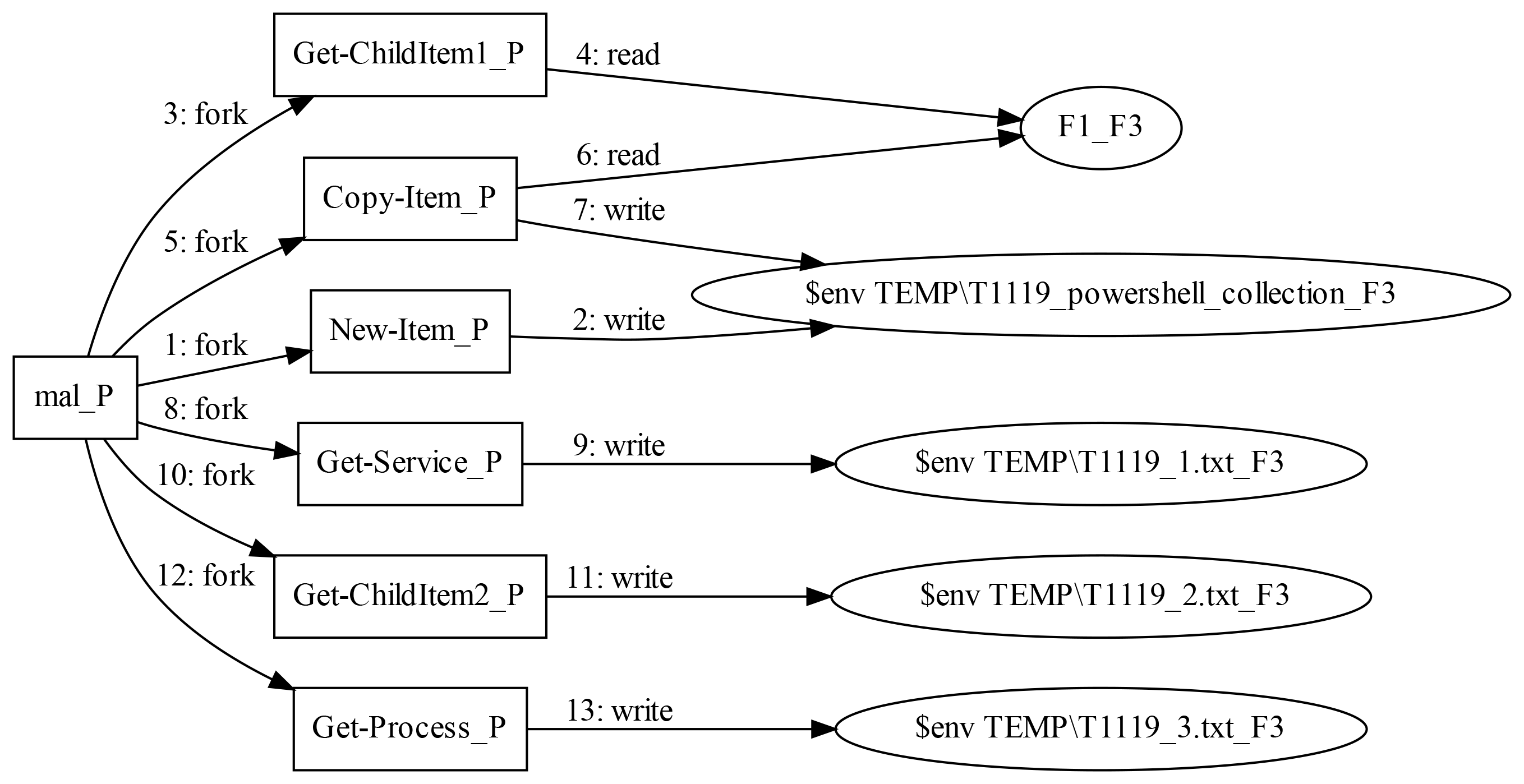}
    \end{minipage}
}
\subfigure[T1003.002]
{
    \begin{minipage}[b]{.45\linewidth}
        \includegraphics[scale=0.4]{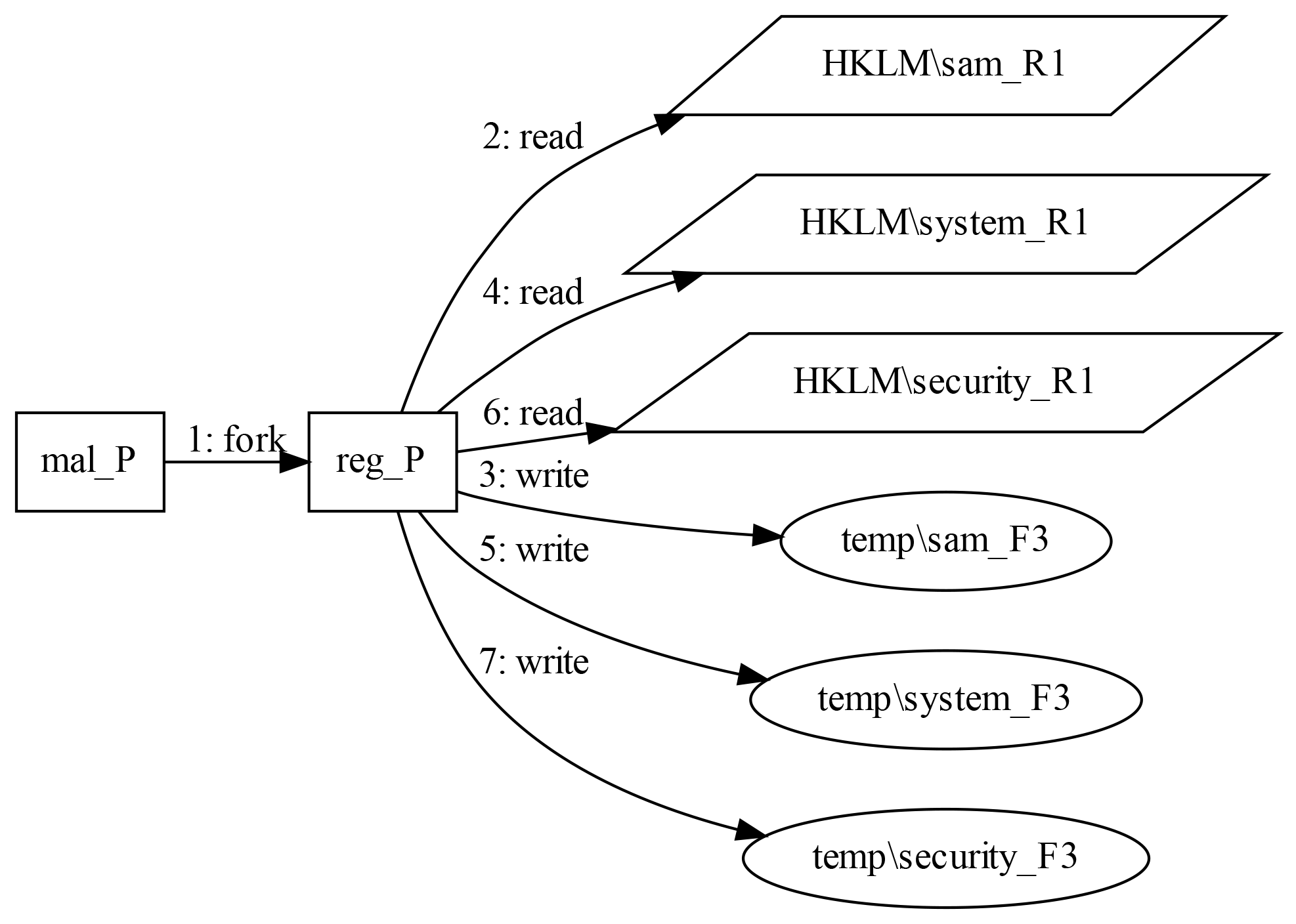}
    \end{minipage}
}
\subfigure[T1083]
{
    \begin{minipage}[b]{.45\linewidth}
        \includegraphics[scale=0.3]{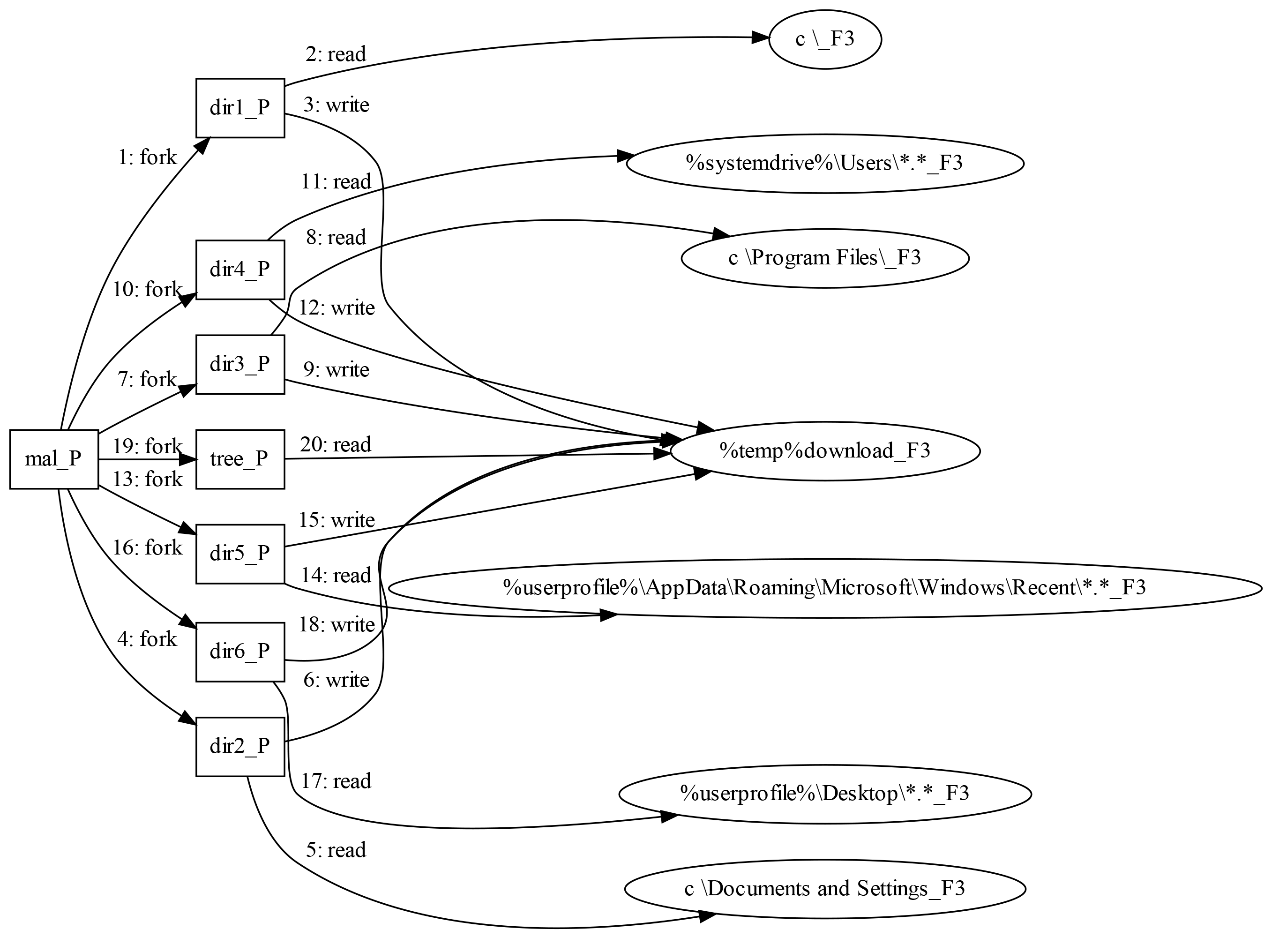}
    \end{minipage}
}
\subfigure[T1036.005]
{
    \begin{minipage}[b]{.45\linewidth}
        \includegraphics[scale=0.5]{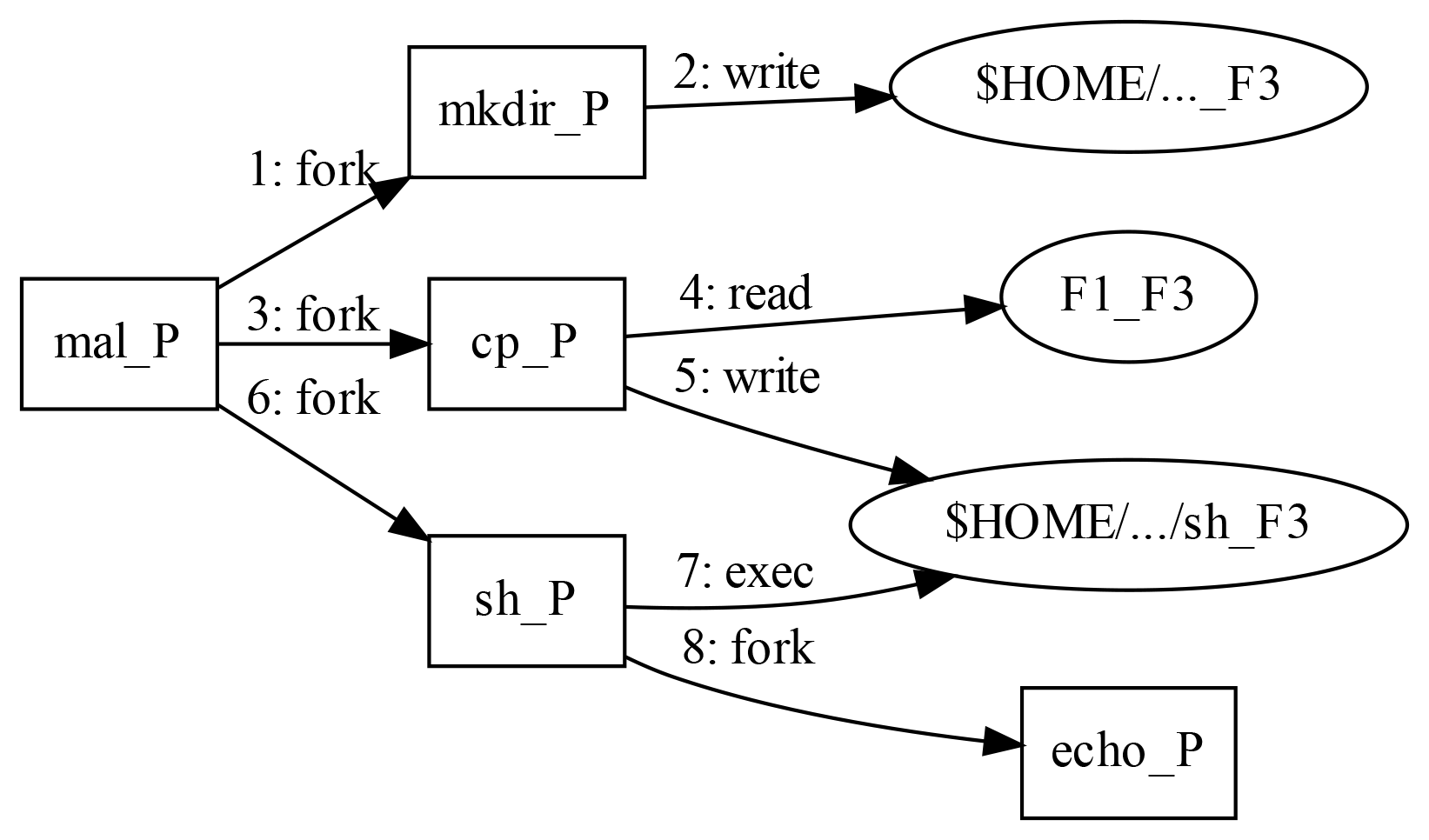}
    \end{minipage}
}
\fbox{
  \parbox[l]{\textwidth}{
     \scriptsize(A) Adversary may use automated techniques for collecting internal data, such as file type, location, or name at specific time intervals.\\
    \scriptsize(B) Adversaries may extract credential material from the SAM database, which contains local accounts for the host.\\
    \scriptsize(C) Adversaries may search files and directories for certain information within a file system to shape follow-on behaviors.\\
    \scriptsize(D) Adversaries may match or approximate the name or location of legitimate files when naming/placing them, for the sake of evading defenses and observation.  
  }
}
\setlength{\abovecaptionskip}{0.cm}
\caption{The examples of manually constructed ATG.}
\label{fig:technique_template_schematic}
\end{figure*}

\section{ATG Examples}
\label{sec:appendix_atgexam}
We manually constructed the ATG for each atomic test by reviewing the descriptions of all the atomic red team techniques and the command lines of the test cases. As shown in Figure~\ref{fig:technique_template_schematic}, we show 4 ATGs along with a brief description of the technique goals.

\section{Approach of evaluating graph forecast model}
\label{sec:appendix_approach_evalGFM}
As described in Section~\ref{sec:6_5} and Figure~\ref{fig:Generation_Model_Evaluation_Method_Design}, we designed an experimental approach to evaluate the effectiveness of the graph prediction model. The red dashed line indicates the removed nodes and dependencies, the blue dashed line indicates the forecast nodes and dependencies, and the shading indicates the technique-level interpretation. The technique-level interpretation results for the original ASG (sub-graph A) are: T1566.001, T1204.002, T1574, for the broken ASG (sub-graph B) are: T1566.001, T1204.002, for the AFG (sub-graph C) are: T1566.001, T1204.002, T1574.

\begin{figure*}[bp]
% \caption{The example of the EFI evaluation experiment on graph forecast and interpretation. Where sub-graph A indicates the original ASG, sub-graph B indicates the broken ASG, and sub-graph C indicates the AFG. The red dashed line indicates the removed nodes and dependencies, the green dashed line indicates the forecast nodes and dependencies, and the shading indicates the technique-level interpretation.}
% \label{fig:Generation Model Evaluation Method Design}
\centering
\subfigure[Original ASG]
{
    % \begin{minipage}[One-hop Edge]{1.0\linewidth}
        \includegraphics[scale=0.6]{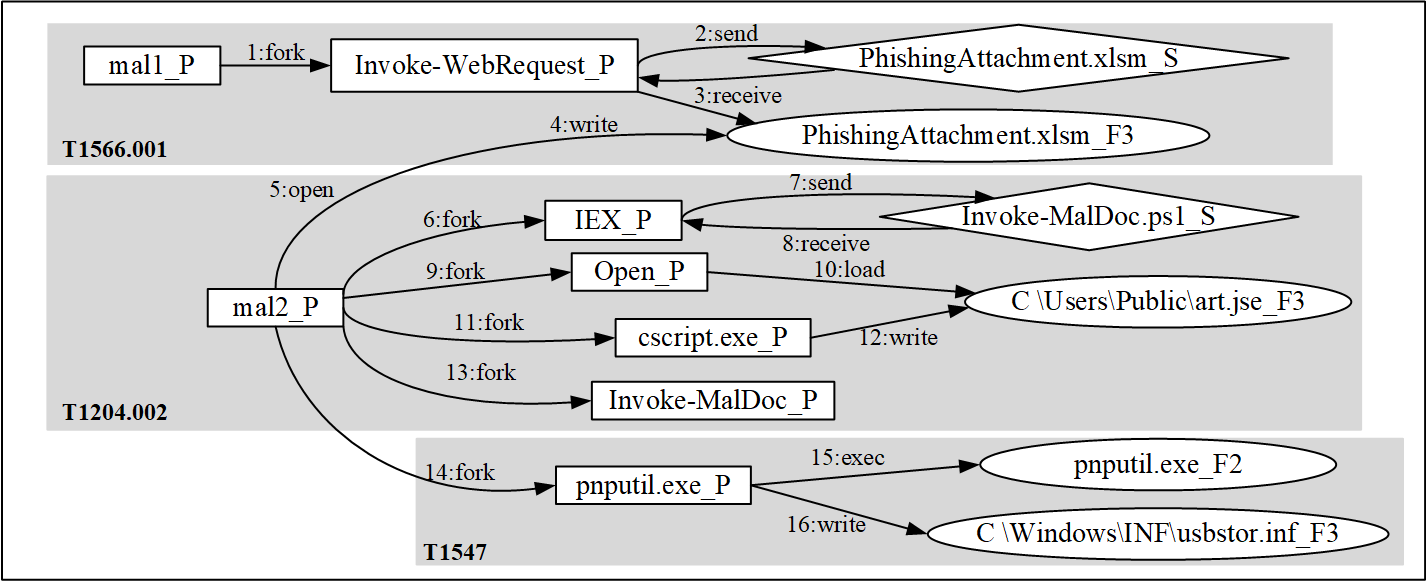}
    % \end{minipage}
    \label{fig:Generation_Model_Evaluation_Method_Design_a}
}

\subfigure[Broken ASG]
{
    % \begin{minipage}[Multi-hop Edge]{1.0\linewidth}
        \includegraphics[scale=0.6]{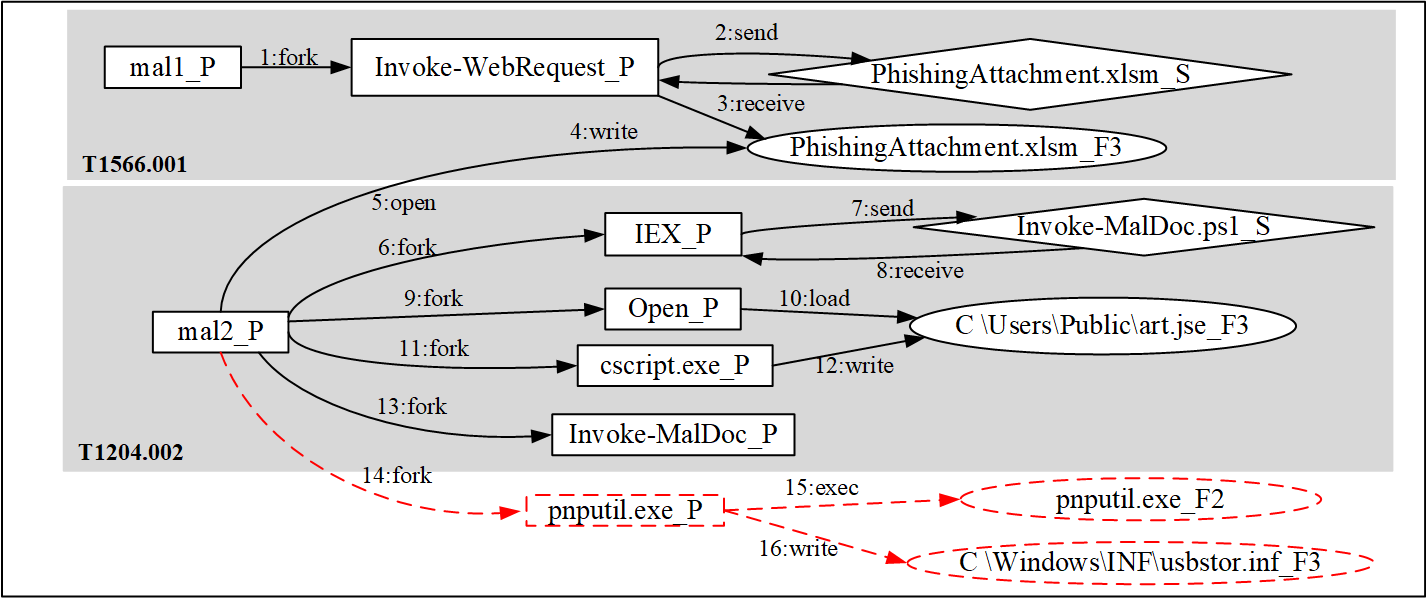}
    % \end{minipage}
    \label{fig:Generation_Model_Evaluation_Method_Design_b}
}

\subfigure[AFG]
{
    % \begin{minipage}[Multi-hop Edge]{1.0\linewidth}
        \includegraphics[scale=0.6]{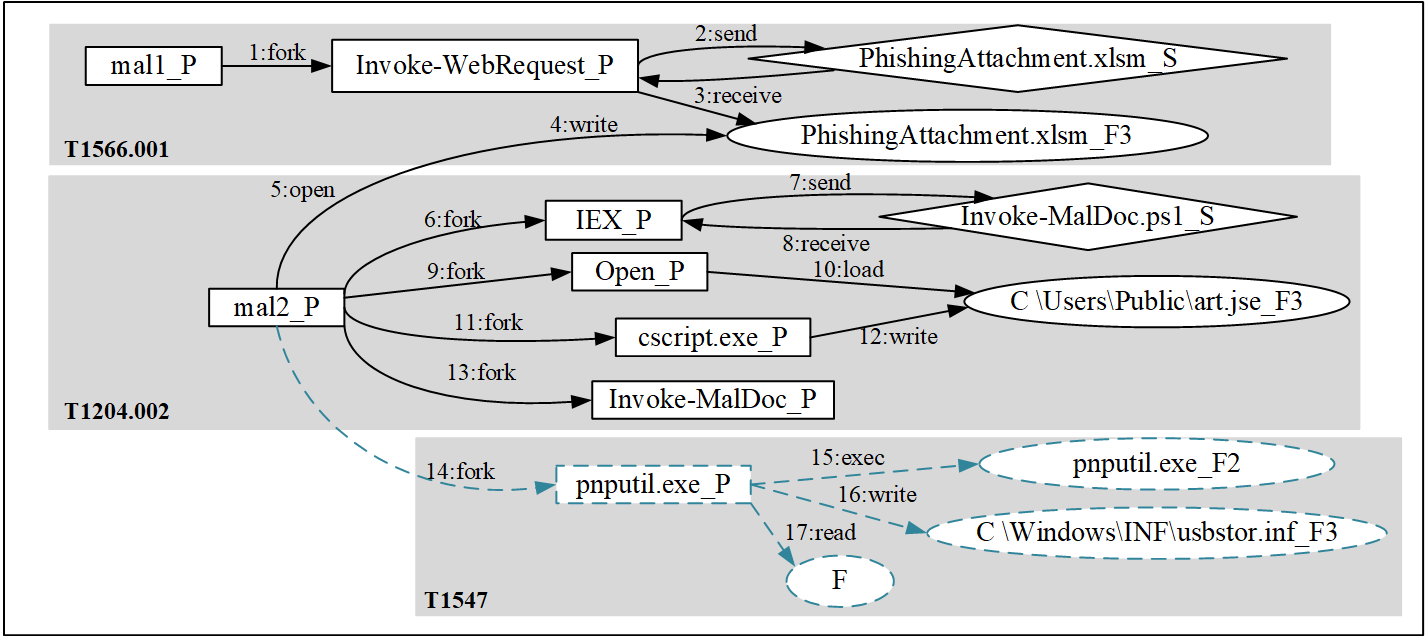}
    % \end{minipage}
    \label{fig:Generation_Model_Evaluation_Method_Design_c}
}
\setlength{\abovecaptionskip}{0.cm}
\caption{The experimental results of graph forecast and interpretation. Figure~\ref{fig:Generation_Model_Evaluation_Method_Design_a} indicates the original ASG, Figure~\ref{fig:Generation_Model_Evaluation_Method_Design_b} indicates the broken ASG, and Figure~\ref{fig:Generation_Model_Evaluation_Method_Design_c} indicates the AFG. The red dashed line indicates the removed nodes and dependencies, the blue dashed line indicates the forecast nodes and dependencies, and the dash area indicates the technique-level interpretation.}
\label{fig:Generation_Model_Evaluation_Method_Design}
\end{figure*}

\end{document}